\documentclass[11pt]{article}  
\usepackage{graphicx}
\usepackage[a4paper,left=3cm,right=3.cm, bottom=2.cm, top=2.0cm]{geometry}
\usepackage{amssymb}
\usepackage{amsmath}
\usepackage{bm}
\usepackage[margin=1pt,font=small,labelfont=bf]{caption}
\usepackage[dvipsnames]{xcolor}
\definecolor{citecolor}{RGB}{128,0,32}
\definecolor{headcolor}{RGB}{128,128,128}
\usepackage[unicode,hyperfootnotes=false,breaklinks=true,colorlinks=true,allcolors=citecolor]{hyperref}
\usepackage{setspace}
\usepackage{fancyhdr}
\usepackage{titlesec}
\usepackage{algorithm}
\usepackage{algpseudocode}
\usepackage{etoolbox}
\newcommand{\degree}{\ensuremath{^\circ}}
\usepackage{times}
\usepackage{tabularx}
\usepackage{datetime} 
\usepackage{lineno}
\usepackage{etoolbox}
\patchcmd{\linenumber}{\hb@xt@}{\hbox}{}{}
\usepackage[ giveninits=true, style=authoryear, backend=bibtex, maxcitenames=2, maxbibnames=99,
  hyperref=true, backref=false, sorting=nyt, uniquename=init, autolang=hyphen,
  dashed=false]{biblatex}

\AtEveryCite{\it \color{citecolor}} \AtEveryBibitem{\clearfield{month}}
\AtEveryBibitem{\clearfield{issn}}
\renewbibmacro{in:}{}
\DeclareGraphicsExtensions{.jpg} \graphicspath{{./jpg/}}
\newdateformat{usvardate}{\shortmonthname[\THEMONTH]. \THEDAY, \THEYEAR}

\fancypagestyle{custom}
{      
\fancyhead[L]{}
\fancyhead[C]{\textcolor{headcolor}{\scriptsize to appear in \textit{Journal of
Geophysical Research: Solid Earth}}}
\fancyhead[R]{}
\fancyfoot[R]{\textcolor{gray}{\thepage}}
\fancyfoot[L]{\scriptsize \url{https://doi.org/10.1029/2025JB032915}}
\fancyfoot[C]{}
}
\renewcommand\thesection{\arabic{section}} \titlespacing\section{0pt}{12pt plus 4pt minus 2pt}{0pt
plus 2pt minus 2pt}
\begin{document}

\pagestyle{custom}
\begin{center}
\LARGE {\bf Fault volume digital twin to reproduce the full slip spectrum, scaling and statistical
laws \\[12pt]}

\normalsize
M. Almakari$^{1,\dagger}$, N. Kheirdast$^{1,\dagger}$, C. Villafuerte$^{1,*}$, M. Y. Thomas$^{2}$,
P. Dubernet$^{1}$, J. Cheng$^{1,\S}$, A. Gupta$^{1}$, P. Romanet$^{3,4}$, S. Chaillat$^{5}$, H. S.
Bhat$^{1,\P}$ \\[12pt]

\begin{enumerate}
	\scriptsize
	\setlength\itemsep{-5pt}
	\item Laboratoire de Géologie, Ecole Normale Superieure, CNRS-UMR 8538, PSL Research University,
	Paris, France\\
	\item Université de Rennes, CNRS, Géosciences Rennes, CNRS-UMR 6118, Rennes
	\item Department of Earth Sciences, La Sapienza University of Rome, Piazzale Aldo Moro 5, 00185
	Roma, Italy
	\item Université Côte d`Azur, CNRS, IRD, Observatoire de la Côte d'Azur, Géoazur,
	Sophia-Antipolis, 06560 Valbonne, France
	\item Laboratoire POEMS, CNRS-INRIA-ENSTA Paris, Institut Polytechnique de Paris
\end{enumerate}

\let\thefootnote\relax\footnotetext{$\dagger$ M. Almakari and N. Kheirdast contributed equally to this work. 
* Currently at Instituto de Geofisica, Universidad Nacional Autonoma de México. 
$\S$ Currently at Division of Geological and Planetary Sciences, California Institute of Technology. 
$\P$ Corresponding author: \texttt{harshasbhat@gmail.com}}

\end{center}

\section*{CRediT}

\scriptsize

\underline{M. Almakari:} Software, Investigation, Writing -- original draft,
\underline{N. Kheirdast:} Software, Investigation, Writing -- original draft, Writing -- review \& editing,
\underline{C. Villafuerte:} Software, Investigation, Writing -- review \& editing,
\underline{M. Y. Thomas:} Methodology, Writing -- review \& editing, Supervision,
\underline{P. Dubernet:} Software,
\underline{J. Cheng:} Investigation, Writing -- review \& editing,
\underline{A. Gupta:} Writing -- review \& editing,
\underline{P. Romanet:} Software, Writing -- review \& editing,
\underline{S. Chaillat:} Software, Writing -- review \& editing,
\underline{H. S. Bhat:} Conceptualization, Methodology, Software, Investigation, Writing -- review \& editing, Supervision, Funding acquisition

\section*{Abstract}
\small
Seismological and geodetic observations of fault zones reveal diverse slip dynamics, scaling, and
statistical laws. Existing mechanisms explain some but not all of these behaviors. We show that
incorporating an off-fault damage zone—characterized by distributed fractures surrounding a main
fault—can reproduce many key features observed in seismic and geodetic data. We model a 2D shear
fault zone in which off-fault cracks follow power-law size and density distributions, and are
oriented either optimally or parallel to the main fault. All fractures follow rate-and-state
friction with parameters enabling slip instabilities. We do not introduce spatial heterogeneities in
frictional properties. Using quasi-dynamic boundary integral simulations accelerated by hierarchical
matrices, we simulate slip dynamics and analyze events produced both on and off the main fault.
Despite spatially uniform frictional properties, we observe a natural continuum from slow to fast
ruptures, as seen in nature. Our simulations reproduce the Omori law, inverse Omori law,
Gutenberg-Richter scaling, and moment-duration scaling. We observe seismicity localizing toward the
main fault before nucleation of main-fault events. During slow slip events, off-fault seismicity
migrates in patterns resembling fluid diffusion fronts, despite the absence of fluids. We show that
tremors, Very Low Frequency Earthquakes, Low Frequency Earthquakes, Slow Slip Events, and
earthquakes can all emerge naturally within this fault volume framework, making it an ideal digital
twin for testing hypotheses, performing ground-truth inversions, and probing mechanical properties
inaccessible with natural observations.

\section*{Plain Language Summary}
Earthquake faults exhibit complex behavior ranging from slow creeping movements to fast destructive
ruptures, but the physical mechanisms underlying this remain unclear. We investigate
whether the geometric arrangement of smaller fractures around a main fault—called the ``damage
zone''—can explain these diverse slip behaviors. We develop simulations that incorporate
both a rough main fault and surrounding networks of fractures, all governed by the same frictional
properties. We find that geometric complexity alone reproduces the full spectrum of fault slip
observed in nature—from barely detectable slow slip events to regular earthquakes—without requiring
spatially varying friction. Our simulations naturally generate realistic
aftershock sequences following Omori's law, earthquake size-frequency distributions matching the
Gutenberg-Richter relation, spatiotemporal clustering of seismicity, and complex source time
functions resembling tectonic tremor. This fault volume model serves as a digital twin that can
generate physically consistent synthetic earthquake catalogs for training machine learning
algorithms, testing how faults respond to external perturbations like tidal stresses or fluid
injection, and developing improved methods for interpreting seismic and geodetic observations. Our
results demonstrate that the network of small fractures surrounding major faults provides a
fundamental organizing principle for fault dynamics, offering a geometric foundation upon which
frictional and hydraulic effects can be evaluated.

\section*{Key Points}
\begin{itemize}
	\item A fault volume model reproduces the full spectrum of slip behaviors—from slow slip events
	to fast earthquakes—using geometric complexity alone.
	\item Damage zone fractures surrounding a rough fault generate realistic aftershock sequences,
	foreshock activity, and spatiotemporal clustering without imposed frictional heterogeneity.
	\item The model naturally recovers empirical scaling laws including Gutenberg-Richter
	statistics, Omori decay, moment-duration scaling, and source-time-function complexity.
	\item Geometric complexity provides a foundational baseline for fault slip dynamics, enabling a
	digital twin framework for testing physical mechanisms and developing inverse methods.
\end{itemize}
\normalsize

\section{Introduction}

In the brittle upper crust, fault zones display a large variety of slip dynamics and moment release.
Until the discovery of slow slip events, SSEs, \parencite{Hirose1999, Dragert2001} and tremor
\parencite{Obara2002}, faults were thought to either remain locked during the interseismic period or
continually creep—the former leading to stick-slip-like release of stored strain energy
\parencite{brace1966b}, the latter to continuous strain release \parencite{steinbrugge1960}. 
Modern seismological and geodetic observations have significantly advanced our understanding of
fault slip behavior, revealing a continuum of deformation modes spanning from regular fast
earthquakes to a diverse family of slow earthquakes and/or steady creep. Slow earthquakes include
``seismic'' members such as low-frequency earthquakes (LFEs), tectonic tremor, and very
low-frequency earthquakes (VLFEs), as well as ``geodetic'' members—slow slip events (SSEs)—typically
classified as short-term (days to weeks) or long-term (months to years) \parencite{Nishikawa2023}.
These phenomena are increasingly recognized as different manifestations of a shared underlying
physical process \parencite{Beroza2011, Ide2023}. 
SSEs, characterized by the gradual release of tectonic stress, have been observed in numerous
subduction zones, including Cascadia, Central Ecuador, Guerrero, Hikurangi, Northern Chile, and
southwest Japan, as well as along continental plate boundaries such as the San Andreas fault system
in California, North Anatolian fault system, Haiyuan fault \parencite{Lowry2001, Dragert2001,
Rogers2003, Douglas2005, jolivet2013, Valle2013, Ruiz2014, rousset2016, Shelly2017, Michel2019,
DalZilio2020}. These events often co-occur with LFEs and tremor, forming Episodic Tremor and Slip
(ETS) sequences, particularly well documented in Cascadia and Nankai \parencite{Rogers2003,
Michel2019}. 
Since their discovery, observations and models have emphasized the potential role of SSEs in the
earthquake cycle. Occurring near or within the seismogenic zone, SSEs are thought to influence the
initiation or modulation of large earthquakes \parencite{Rogers2003, Segall2012, Obara2016,
CruzAtienza2021}.
Moreover, aseismic slip phenomena are not limited to traditionally aseismic regions. Increasingly,
observations show that slow and fast slip events can coexist within the seismogenic zone itself
\parencite{Schwartz2007, Ito2013, Ruiz2014, Thomas2017}. Aseismic slip has notably been observed
prior to several major earthquakes, including the 2011 Tōhoku-Oki \parencite{Ito2013}, 2014 Iquique
\parencite{Ruiz2014}, and 2017 Valparaíso earthquakes \parencite{Ruiz2017, Caballero2021}.
Postseismic aseismic slip in seismic patches has also been documented in several
subduction/collision zones \parencite{Johnson2012, Thomas2017, Villafuerte2025}, further
highlighting the complex interaction between seismic and aseismic processes.

Although slip dynamics may appear highly complex across timescales, robust empirical laws continue
to emerge from seismological observations of fast events. The Gutenberg-Richter law describes the
magnitude-frequency distribution of earthquakes, following a power law with a b-value near 1
\parencite{Gutenberg1942}. The Omori law characterizes the decay rate of aftershocks following a
mainshock \parencite{Utsu1995}, while an inverse Omori law describes the acceleration of foreshock
activity before major events, first noted by \textcite{Papazachos1973}. For fast ruptures, seismic
moment scales cubically with rupture duration \parencite{Ide2008, Ide2023}. For slow ruptures,
debate remains between a linear scaling \parencite{Ide2008} and a cubic one \parencite{Gomberg2016,
Michel2019}. More recently, \textcite{Kato2020} reported a phenomenon of localization followed by
delocalization of deformation before and after major earthquakes, respectively—a pattern observed in
southern California for several large events, including the 1992 $M_w$ 7.3 Landers, 1999 $M_w$ 7.1
Hector Mine, and 2019 $M_w$ 7.1 Ridgecrest earthquakes \parencite{BenZion2020}. Furthermore,
seismicity has been observed to migrate during slow-slip events, with spatiotemporal features
resembling a diffusion front, suggesting that fluid diffusion may govern these dynamics
\parencite[e.g.][]{Danr2024}.

The wide spectrum of observed slip behaviors has prompted investigation into various physical
mechanisms, with fault friction heterogeneity emerging as the most commonly invoked
\parencite{brace1966b, scholz2019}. Laboratory experiments have consistently shown that frictional
properties are central to the transition from stable to unstable slip, particularly near the
brittle-ductile transition zone \parencite[e.g.][]{Leeman2016, Tinti2016, Scuderi2016, Scuderi2017,
Leeman2018, Sirorattanakul2024, Pignalberi2024, Yuan2024, Meyer2024, SalazarVsquez2024}. Numerical
simulations incorporating variable constitutive friction parameters within rate-and-state frameworks
have successfully reproduced a spectrum of slip modes, including slow ruptures
\parencite{Yoshida2003, Liu2005, Barbot2019, Nie2021}. Modeling and experimental studies indicate
that frictional heterogeneity not only modulates slip dynamics and triggering \parencite{Aochi2009,
Dublanchet2013} but also governs complex precursory processes that may culminate in a mainshock
\parencite{Kato1997, Ariyoshi2012, Dublanchet2018, Gounon2022, Wang2023b}. Other mechanisms have
also been proposed, including fluid-related effects such as dilatant strengthening
\parencite{segall1995, Segall2010, Liu2010}, spatiotemporal variations in permeability and pore
fluid pressure \parencite{Skarbek2016, CruzAtienza2018, Zhu2020, PerezSilva2023, Ozawa2024}, elastic
or poroelastic bimaterial effects \parencite{Heimisson2019, Abdelmeguid2022}, brittle patches
embedded in ductile matrices \parencite{ando2012}, and transient stress perturbations due to nearby
earthquakes \parencite{liu2007} or failed nucleation processes \parencite{rubin2008}. Notably, the
earliest model of slow slip events was proposed by \textcite{perfettini2001}, who showed that a
spring-block system obeying rate-and-state friction and subjected to normal traction perturbations
could, under certain conditions, exhibit ``aseismic stick-slip'' behavior—prior to the observational
discovery of SSEs in Cascadia and the Bungo Channel. All of the above mechanisms primarily focus on
modeling the fault as a single frictional surface.

However, faults are far more complex than simple, narrow shear zones idealized as frictional
surfaces. Geological observations show that, surrounding the fine-grained fault core—where most slip
localizes—is a broader damage zone composed of pervasively fractured rock with an intricate
three-dimensional geometry across multiple length scales \parencite{sibson1977, chester1993,
biegel2004}. 
This hierarchical geometric complexity ranges from tens of kilometers \parencite[see, for
e.g.,][]{Fletcher2014} down to the millimetric scale \parencite{Sowers1994, Mitchell2009,
Fagereng2010}. Figure 1 of \textcite{Okubo2019} illustrates this hierarchical structure of fault
systems over a wide range of length scales. Fault slip typically occurs on a localized plane—the
fault core—which lies within a damage zone that generally extends several hundreds of meters in
width \parencite{Chester1986, sibson1986, Power1987, benzion2003a, Sibson2003, Savage2011,
Ostermeijer2020, RodriguezPadilla2022, Liu2025}. Various studies have shown that fault geometry—such
as roughness \parencite{Candela2011, Cattania2021, Ozawa2021}, bends \parencite{King1985,
ritz_influence_2015,Romanet2020, Ozawa2023}, branches \parencite{Aochi2000, Kame2003a, Kame2003,
Oglesby2003a, Bhat2004, Bhat2007b, Marschall2024,templeton2010}, and step-overs \parencite{Oglesby2005, Biasi2016,
RodriguezPadilla2024}, play an important role in the dynamics of earthquake ruptures.
Recent work has also underlined the key role of active faults networks \parencite{Im2023,
Cheng2025}. Large earthquakes ($M_w > 7$) often exhibit complex ruptures involving multiple faults
\parencite{Stein2024}. For example, the 1992 Landers earthquake involved the activation of multiple
faults during a single seismic event \parencite{Sowers1994, Fliss2005}, and the 2016 Kaikōura
earthquake involved at least 15 faults \parencite{Klinger2018}. More recently, fault system geometry
was found to significantly influence the slip distribution of the $M_w$ 7.1 Ridgecrest earthquake
\parencite{Nevitt2023} and the 2023 $M_w$ 7.8  Kahramanmaraş, Turkey, earthquake
\parencite{Chen2024NatCom,Yao2025}. Furthermore, recent work by \textcite{Lee2024} suggests that fault slip
stability may be controlled by the orientation and complexity of surrounding fault networks, with
complex fault systems associated with locked segments that promote stick-slip behavior, whereas
simpler geometries favor stable creep.

Several studies have also focused on understanding the role of newly created fracture - or dynamic
damage - in the propagation of a single dynamic rupture \parencite[][among others]{xu2015,bhat2012}. This damage can significantly affect the
high-frequency radiation generated during rupture propagation \parencite{Thomas2018, Okubo2019,
Marty2019, Okubo2020}. Recent findings highlight radiation from such multifracture structures in
subduction zones \parencite{Chalumeau2024}, and the anisotropy imposed by the preferred orientation
of off-fault fractures \parencite{Huang2025}. Moreover, as a dynamic rupture propagates, the energy
dissipated by off-fault fracturing processes (creation of new fractures or reactivation of
preexisting ones) becomes critical to consider \parencite{Okubo2019}. The fracture energy dissipated
within the off-fault volume can be substantial—often comparable to the fracture energy on the main
fault itself \parencite{Andrews2005, Okubo2019, Okubo2020}. Measurements of off-fault inelastic
deformation using radar and optical imagery, when compared with aftershock sequences, indicate that
fault systems exhibiting greater off-fault damage tend to produce relatively higher numbers of
aftershocks \parencite{Milliner2025}. Additionally, dynamic damage has been shown to play a crucial
role in controlling the transition to supershear rupture \parencite{Jara2021}. 

Thus, while fault slip behavior is strongly influenced by the properties of the interface itself,
the overall mechanical behavior of faults is equally shaped by the structural complexity of the
entire fault zone. Its physical properties evolve over timescales ranging from seconds to millions
of years, accommodating displacements from millimeters to tens of kilometers. In particular, dynamic
ruptures can induce significant changes in both on-fault and off-fault mechanical properties. These
changes, in turn, influence rupture nucleation, propagation, timing, seismic wave radiation, and
postseismic deformation. This broader structure alters the rheology of both the fault core and the
surrounding rock, introduces complex interactions, and thereby affects the style and dynamics of
fault slip \parencite[e.g.][] {Andrews2005, collettini2009, niemeijer2010, thomas2014b,
Faulkner2006, dor2006a, dor2006b, Mitchell2009, bhat2010a, biegel2010, bhat2012, Okubo2019, Okubo2020}.
Understanding such processes inevitably calls for a comprehensive approach that considers the fault
zone as a whole, rather than focusing solely on the fault core.

Keeping this in mind, recent studies increasingly focus on understanding the role of complex and
irregular geometries throughout the full seismic cycle. \textcite{Romanet2018} incorporated
geometrical complexity in numerical models and found that stress interactions between parallel
faults can naturally generate slow slip events. Similarly, \textcite{Cattania2021} showed that
foreshock sequences can emerge spontaneously when rough fault surfaces are considered. In a related
study, \textcite{Ozawa2021} modeled a fault network with subsidiary fractures surrounding a primary
rough fault and demonstrated the spontaneous emergence of an Omori-like aftershock decay
\parencite{Utsu1995}. Incorporating the realistic geometry of northern Cascadia into numerical
models enabled the generation of slow slip events that closely match GPS observations in the region
\parencite{Li2016}. Although an increasing number of studies \parencite{Yin2023, Im2024, Peng2025}
explore the effects of geometry over multiple seismic cycles, no current model reproduces the full
range of observed slip dynamics.


We aim to investigate the role of this realistic fault geometry and its associated off-fault
damage—hereafter referred to as the fault volume—in governing slip event dynamics. We propose a
simplified model of a 2D fault volume consisting of a main self-similar rough fault surrounded by a
hierarchy of off-fault slip planes. All fractures are frictionally homogeneous (rate-weakening) and
capable of dynamic slip. To enable simulation of this geometrically complex system, we develop a
fast quasi-dynamic earthquake cycle model accelerated using hierarchical matrices. Hierarchical
matrices compress the stress interaction matrix by approximating distant interactions with low-rank
blocks, enabling simulations of large, complex systems within reasonable computational time (see
Supporting Information for more details). Remarkably, this purely geometric complexity is sufficient
to reproduce the full spectrum of observed slip behaviors. The model can generate a continuum of
slip events and recover all major empirical scaling laws associated with the seismic cycle. This
establishes the model framework as a ``digital twin'' of a fault zone where hypotheses about fault
slip mechanisms can be tested systematically, inversion techniques can be validated against known
ground-truth, and fundamental mechanical properties—such as stress evolution, energy partitioning,
and damage zone dynamics—can be probed directly in ways that are impossible with natural observations. 

In the following section, we introduce the fault volume model, stress loading, and initial
conditions used in our study. We then outline the methodology for event detection, distinguishing
between fast and slow ruptures, and describe our approach for constructing the synthetic earthquake
catalog. In the results section, we begin with a case study illustrating the spatiotemporal
evolution of seismic cycles and the emergence of complex slip dynamics. We analyze the moment rate
functions of events across scales, revealing distinct characteristics between slow and fast
ruptures. We then demonstrate the model's ability to reproduce key empirical laws, including the
Gutenberg-Richter relation, moment-rupture area scaling, moment-duration scaling, and both Omori and
inverse Omori laws. Additionally, we highlight the model's capacity to replicate the
localization-delocalization transition of deformation and the apparent diffusion-like migration of
seismicity during slow slip events. We conclude by discussing the model's implications and
limitations, identifying testable hypotheses, and evaluating its potential as a digital twin
framework for fault systems.

\section{Fault Volume Model}
 
\subsection{Overview of the Fault Volume Model}

This study focuses primarily on the impact of the geometry and architecture of a fault zone on slip
dynamics. Hence, this fault volume model is a simplified representation of a fault and its damage
zone.  We begin by defining the elastic properties of the medium, its loading and initial
conditions. All length scales are non-dimensionalized using frictional length scales (either the
nucleation length or the cohesive zone size) computed using the initial stress state of the medium.
We remark that these frictional length scales (see Supporting Information for more details) are
merely approximations, as they are based on assumptions of a single planar fault undergoing no
change in normal and shear traction over time and always at steady state. We then establish the
frictional properties of the main fault and set its length to four times the nucleation length.
Next, we define a damage zone with a width, $W$, five times the cohesive zone size of the main
fault. Within this damage zone, we assume a hierarchical distribution of off-fault fractures. The
density of these fractures decreases according to a power law with a fixed exponent and maximum
value. The size distribution also follows a power law with a fixed exponent, where the smallest
length scale is approximately 1m and the largest is on the order of the main fault length. We select
the orientation distribution of these fractures from one of four possible orientations with respect
to the main fault and principal stress directions (Section~\ref{geometeryfaultvolume}). We assume
each off-fault fracture is twice its nucleation length to determine the critical slip distance,
$d_c$, for each fracture. This ensures $d_c$ scales automatically with fault length as described
above. Finally, we statistically sample five times from the distributions (length, orientation, and
density) to generate five samples of the fault volume model for a given orientation of the off-fault
fractures and its rate and state friction properties. We thus generate a total of 40 different fault
volume models for this study each of them including around 700-1000 faults. In the following
sections we describe each of the steps in more detail.

\subsection{Geometry of the Main Fault and its Damage Zone}\label{geometeryfaultvolume}

\begin{figure}[t!]
	\centering
	\includegraphics[width=\textwidth]{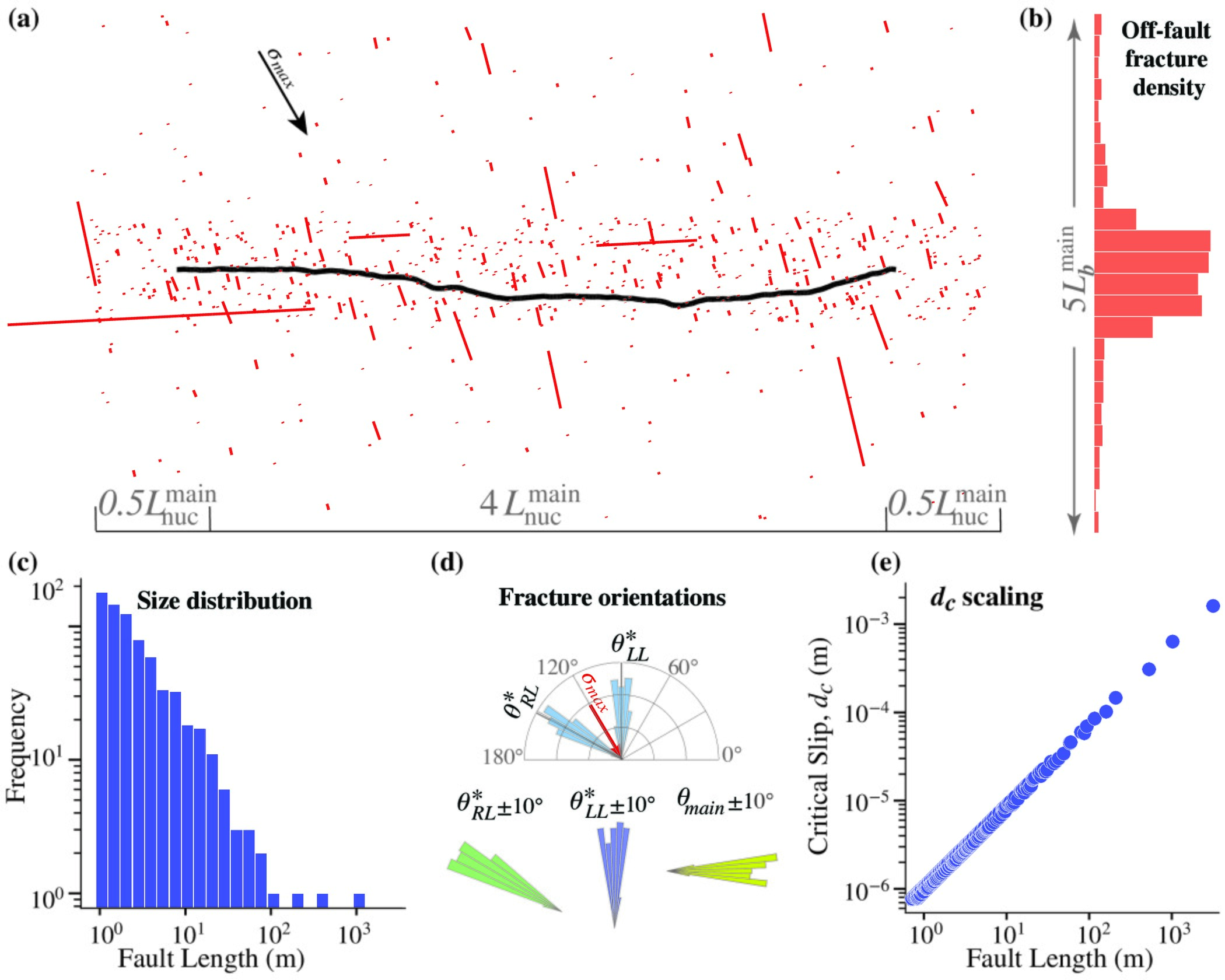}
	\caption{Fault volume geometry of a case study: a) sketch of fault volume geometry (not to
		scale): the main fault in black and the off-fault fractures in red  b) off-fault fracture
		density perpendicular to the main fault (linear scale) c) Frequency of length distribution
		of the faults, including the main fault d) rose diagram of off-fault fracture orientations
		and principal stress direction e) critical slip distance of all fractures with respect to
		their length. RL and LL stands for right and left lateral slip respectively.}
	\label{geometry}
\end{figure}

The fault volume in the model comprises a main fault exhibiting right-lateral movement in an
isotropic linear elastic medium ($\mu = 30$ GPa, $\nu = 0.25$), with fixed frictional properties,
accompanied by off-fault fractures within a damage zone surrounding the main fault. Since geometric
roughness is a characteristic observed at multiple scales in natural faults \parencite{Power1987,
Schmittbuhl1993, Lee1996, Renard2006, Candela2009, Candela2012}, in this fault volume model, the
main fault is characterized by self-similar roughness with $\alpha = 0.002$ \parencite{Dunham2011}.
Its extent is four times its nucleation length $L_{\text{main}} = 4 L_{\text{nuc}}^{\text{main}}$
(Figure~\ref{geometry}a). Note that the nucleation length is just an estimation, calculated for a
single planar infinite fault, following \textcite{Rubin2005,viesca2016a} (See Supporting Information
for more details). We set the damage zone width as $W = 5 L_b^{\text{main}}$ making it proportional
to the main fault's cohesive zone size ($L_b^{\text{main}}$). This choice reflects the fact that the
cohesive zone size governs the decay rate of stress perturbations from the main fault, which in turn
controls the characteristic length scale over which main and off-fault fractures interact
\parencite{Okubo2019}. We also add a fracture damage zone at each end of the main fault whose
fracture density is assumed to be uniformly distributed.

From natural observations, the length distribution of off-fault fractures follows a power law
\parencite{Bonnet2001, benzion2003a}. This modeling choice is grounded in the understanding that
such distributions effectively capture the natural variability of fault lengths. Accordingly, we
consider a power law distribution of fractures length with an exponent of 2 (Figure
\ref{geometry}c). The smallest length scale is set to be around 1 m and the largest being of the
order of the length of the main fault. Each off-fault fracture is set to be twice its nucleation
length, ensuring it is large enough to rupture independently \parencite{Rubin2005,viesca2016a}.

We next set the orientation statistics of the off-fault fractures. We experiment with four different
orientations of fractures. In the first three cases, the fractures are deliberately well oriented to
maximize reactivation and thus interaction with the main fault. We examine favorable orientations
for right-lateral failure, left-lateral failure, and a combination of both (see example in
Figure~\ref{geometry}d). Optimal planes make an angle of $\theta = \pm \left( {\pi}/{4} - {1}/{2}
\tan^{-1} f_0 \right)$ with the direction of maximum principal stress \parencite{Bhat2007a}, where
$f_0$ is the reference friction coefficient. Here $f_0 = 0.6$, so $\theta \approx \pm 30\degree$.
The initial maximum principal stress makes an angle of 120$\degree$ with the horizontal axis. We
thus consider right-lateral fractures (oriented at $\theta = 150\degree \pm 10\degree$),
left-lateral fractures (oriented at $\theta = 90\degree \pm 10\degree$), both right and left-lateral
fractures (conjugate planes at $\theta = [90\degree; 150\degree] \pm 10\degree$). For the final
case, the fractures run parallel to the main fault, as it has been observed that the growth process
of faults reveals that off-fault cracks tend to be oriented at angles of [0\degree–20\degree] to the
trace of major faults \parencite{Perrin2015}. For this case, we consider the fracture orientations
in the range [0\degree ± 10\degree] (Figure~\ref{geometry}d).

Observations from fault zones consistently show a significant decrease in the density of off-fault
cracks with increasing distance from the main fault. This observation has been well documented in
fault zones \parencite{Chester1986, Faulkner2006, Powers2010, RodriguezPadilla2022}. Based on these
observations of damage zones surrounding major faults, we model a power-law decay of fracture
density with distance normal to the main fault, using an exponent of 1 (Figure \ref{geometry}b shows
the sampled distribution). This exponent is higher than values reported in some studies
\parencite{Ostermeijer2020, Savage2011} but lower than the values suggested by
\textcite{Powers2010}. We fix the peak density of off-fault fractures to be around 10 fractures per
meter, which is consistent with observations \parencite{Chester1986, Mitchell2009,
faulkner2011, RodriguezPadilla2022}. We further assume that all of the off-fault fractures are pure
mode II fractures. The main numerical bottleneck in simulating tensile off-fault fractures in an
earthquake cycle model is the lack of a good cohesive law that allows for the healing of the
cohesive strength after rupture. We therefore restrict ourselves to shear fractures that follow
rate-and-state friction law.

\subsection{Frictional Parameters}

Friction is assumed to follow regularized rate-and-state friction, spatially uniform and
rate-weakening on all faults with $f_0 = 0.6$ (See Supporting Information for more details).
Specifically, for the main fault, the ratio $a/b$ is fixed at 0.75 and $d_c$ is set at 2 mm. Here,
$a$ is the direct effect parameter that governs the instantaneous change in friction with a change
in slip rate, $b$ is the evolution effect parameter which controls how friction evolves over time
via changes in the state variable and $d_c$ is the critical slip distance. Since $a/b$ is a crucial
parameter in controlling the frictional length scales (see Supporting Information for more details)
we parameterize the problem by varying this parameter. All off-fault fractures within the model share
the same $a/b$ ratio, with tested values of 0.4 and 0.5. The characteristic slip distance ($d_c$) is
scaled with the length of off-fault fractures (Figure \ref{geometry}e) such that each off-fault
fracture is twice the nucleation length. 

Earthquake sources are conventionally modeled as shear ruptures occurring on preexisting faults
within the seismogenic zone. Nevertheless, these faults are characterized by complex geometric
irregularities and mechanical heterogeneities that manifest across a wide range of spatial scales
throughout the fault zone. Be it in the lab scale or the field scale, the net weakening distance is
a homogenized manifestation of weakening processes occurring at smaller scales
\parencite{ohnaka1999,Gabriel2024}. This approach is also inspired by the observation that the so
called fracture energy ($G'$) scales with slip \parencite{Ohnaka2003, Abercrombie2005}, and slip is
itself related to fracture length, through elasticity. Furthermore, \textcite{Rubin2005} showed that
fracture energy is approximately proportional to $d_c$ within the rate-and-state friction framework.
These observations together justify scaling $d_c$ with fracture length in our model. This also has
the additional advantage of keeping the computational cost reasonable. To keep the values of $d_c$
realistic, we ensure that the smallest fracture is around 1m, the length scale of laboratory
experiments and use laboratory inferred values of $d_c$ \parencite{marone1998, Scuderi2016,
Leeman2018} of the order of several tens of microns.

\subsection{Loading and Initial Conditions}

The medium has an initial prestress state with $\sigma_{11}^0 = -6.46\,\text{MPa}$, $\sigma_{12}^0 =
3.88\,\text{MPa}$, $\sigma_{22}^0 = -10.94\,\text{MPa}$. It is also loaded with uniform far-field
stressing rate, $\dot{\sigma}_{11}^\infty = -0.0064\,\text{Pa\,s}^{-1}$, $\dot{\sigma}_{12}^\infty =
0.0038\,\text{Pa\,s}^{-1}$, $\dot{\sigma}_{22}^\infty = -0.0108\,\text{Pa\,s}^{-1}$. This
corresponds to strain rates $\dot{\varepsilon}^{\infty}_{11} = -4.93 \times 10^{-14} \text{ s}^{-1},
\dot{\varepsilon}^{\infty}_{22} = -1.23 \times 10^{-13} \text{ s}^{-1}$ and
$\dot{\varepsilon}^{\infty}_{12} = 6.33 \times 10^{-14} \text{ s}^{-1}$. The direction of principal
stress for far-field stress rate and initial stress are the same, and is assumed here to make an
angle of 120$\degree$ with the horizontal axis. At each fault, $\tau_0 = \sigma_{ij}^0 n_j s_i$ and
$\sigma_0 = \sigma_{ij}^0 n_j n_i$, where $\tau_0$ and $\sigma_0$ are initial shear and normal
stress on the fault, $\mathbf{n}$ and $\mathbf{s}$ are the unit normal and unit tangent vectors respectively. All
faults are assumed to be initially at steady state, and initial slip rates are considered constant:
$V_{\text{init}} = V_0 = 10^{-9}$ m/s where $V_0$ is the reference slip rate for rate-and-state
friction. The system initiates through the activation of a high slip rate patch located on the main
fault. We neglect the first artificial earthquake on the main fault forced by loading, and only
study the subsequent cycles after the system loses the memory of the initial conditions. To avoid an
unusual buildup of normal traction on the fault, we simply cap the normal traction at 10 MPa in an
elastic-rigidly plastic sense. This needs to be examined in more detail in future work as it is
beyond the scope of this paper. The slip rate is capped at a minimum value of $10^{-20}$ m/s at the
end of each time step to avoid unnecessary numerical artifacts as suggested in the SEAS benchmark
\parencite{Jiang2022}.  

\subsection{Numerical Implementation}\label{numericalimplementation}

We implement the fault volume model using a quasi-dynamic boundary integral method with spatial
convolution accelerated using hierarchical matrices. To assess the sensitivity of our fault volume
model to various parameters, we conducted multiple simulations exploring different fracture
orientations and frictional parameters of off-fault fractures. To ensure robustness of our results,
each off-fault orientation was subjected to five different statistical samplings, guaranteeing a
comprehensive understanding of the system's behavior across various geometries. Furthermore, we
repeated the entire analysis, initially with an $a/b$ ratio of 0.5 for the off-fault fractures, and
subsequently with a new $a/b$ ratio of 0.4, while maintaining homogeneous, rate-weakening friction
conditions. In total, we tested 40 different configurations with various combinations of off-fault
fracture orientations and frictional parameters.

We account for the full elastic interaction between all faults in the system, including the main
fault and all off-fault fractures. To ensure numerical accuracy, we discretize each of the faults
such that at least 5 grid points are within the cohesive zone of each fault. This results in about
30000–50000 unknowns in the system per simulation. Each of the simulations is run until we produce
between 8 and 10 major earthquakes on the main fault, which corresponds to between 2 and 4 million
adaptive time steps. More details on the numerical implementation can be found in Supporting
Information.

\subsection{Event Detection and Catalog Generation}\label{catalog}

\begin{figure}[!ht]
	\centering
	\includegraphics[width=0.9\textwidth]{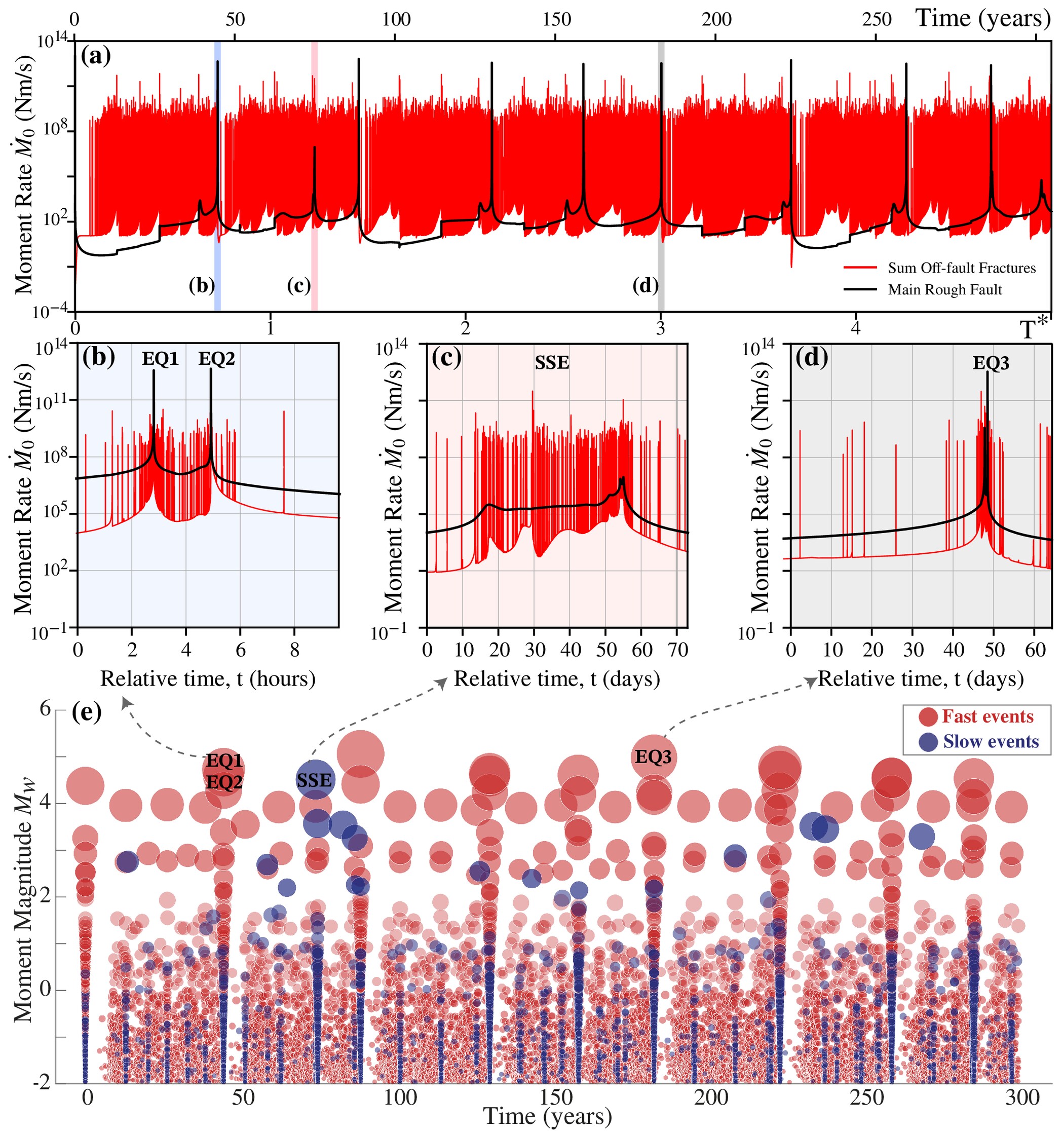}
	\caption{Time series of the moment rate $\dot{M}_0$. The black curve represents the contribution
	coming from the main rough fault, and the red curve represents the summation of the moment rate
	released by all the fractures. Panel a) The entire seismic cycle. Top x-axis is absolute time in
	years, and bottom x-axis is time normalized by recurrence time of earthquakes when considering
	only the main rough fault, without a damage zone ($\sim 62$ years). Panel b), c) and d) show
	time snapshots of the seismic cycle to highlight certain events. Here x-axis is relative time
	for the corresponding time snapshot. The spatiotemporal description of these events are shown in
	Figure \ref{slip-rate-profile}. Panel e) shows the catalog built from the above continuous moment-rate results using slip velocity threholds as described in section \ref{catalog}.} 
	\label{moment-rate}
\end{figure}

The detection and classification of events in the system involve several steps. Events are
identified when the maximum slip rate exceeds predefined thresholds, with categorization into
``fast'' or ``slow'' depending on the specific threshold exceeded: $10^{-3}$ m/s for fast and
$10^{-6}$ m/s for slow. Additional values for the slow rupture threshold ($10^{-7}$, $10^{-8}$, and
$10^{-9}$ m/s) are also tested. The effect of this threshold on the catalog and the scaling laws
will be discussed in the following sections. Subsequently, events that fall within this threshold
range are delineated spatiotemporally through the application of the connected-component labeling
(CCL) algorithm, commonly utilized in image processing. The detailed event detection algorithm is
presented in Supporting Information. This process yields information on rupture length
$L_{\text{rup}}$ and duration $T$, which allows us to estimate an average rupture velocity $v_r =
L_{\text{rup}}/T$. We also estimate for each event the final slip $\delta$ and subsequently the
seismic moment as follows: $M_0 = \mu S \delta$, where $\mu$ is the shear modulus and $S$ is the
rupture area. Since the model is 2D, we use an equivalent rupture surface $S$ defined as $\pi
L_{\text{rup}}^2/4$, assuming a circular rupture. We then estimate the moment magnitude $M_w = 0.67
\log_{10}M_0 - 6.06$ \parencite{Hanks1979}. Finally, we apply a filtering process to the detected
events, eliminating those with fewer than 5 time steps and spanning less than 5 grid points so that
the catalog properties can be accurately computed. An example of the generated catalog is shown in Figure \ref{moment-rate}e.


\section{Influence of the Fault Volume on the Slip Dynamics of the System}
\label{case-study}

\subsection{Seismic Cycles and Slip Dynamics}\label{sec:slip_dynamics} 

In this section, we provide a concise summary of results based on a case study using the fault
geometry shown in Figure~\ref{geometry}.

Figure~\ref{moment-rate}a depicts the time series of the moment-release rates ($\dot{M}_0$), with the
black curve illustrating the contribution originating from the main rough fault. In contrast, the
red curve represents the sum of the moment-release rates of off-fault fractures. In
Figure~\ref{moment-rate}a, in order to measure the impact of the fault volume on slip dynamics, we
normalize time by the recurrence time of earthquakes when considering only a single rough fault,
without a damage zone. Notably, for
the fault volume scenario presented here, we observe eight ruptures, twice as many as in the
single-fault case, over the same time period (Figure~\ref{moment-rate}a).
This substantial difference emphasizes the significant perturbation introduced by the fault volume
on the seismic behavior of the system, resulting in a discernibly shorter recurrence interval.
Importantly, the recurrence time is not constant but rather exhibits variability, giving rise to
intricate seismic and aseismic events. While the main fault experiences seismic cycles, individual off-fault
fractures also undergo their own seismic cycles, but their collective moment release appears
relatively steady throughout the main fault's seismic cycle. We will examine this in more detail in
later sections.

We observe both slow and fast earthquakes on the main fault, in comparison to the dynamics of a
single main fault where only fast ruptures are observed. Figures~\ref{moment-rate}b–d shed light on
the complex seismic behavior driven by the presence of a fault volume. Examples include instances of
multiple ruptures occurring within a brief timeframe, with intervals as short as a few hours between
them (Figure~\ref{moment-rate}b). Additionally, we observe an increase in seismic activity recorded
on the fractures before the rupture and a subsequent decrease afterward. This will be further
discussed in the following sections. Furthermore, Figure~\ref{moment-rate}c reveals the occurrence
of prolonged slow slip events spanning a couple of months. These findings collectively underscore
the complex nature of seismic activity influenced by the presence of a fault volume.


\begin{figure}[!ht]
	\centering
	\includegraphics[width=\textwidth]{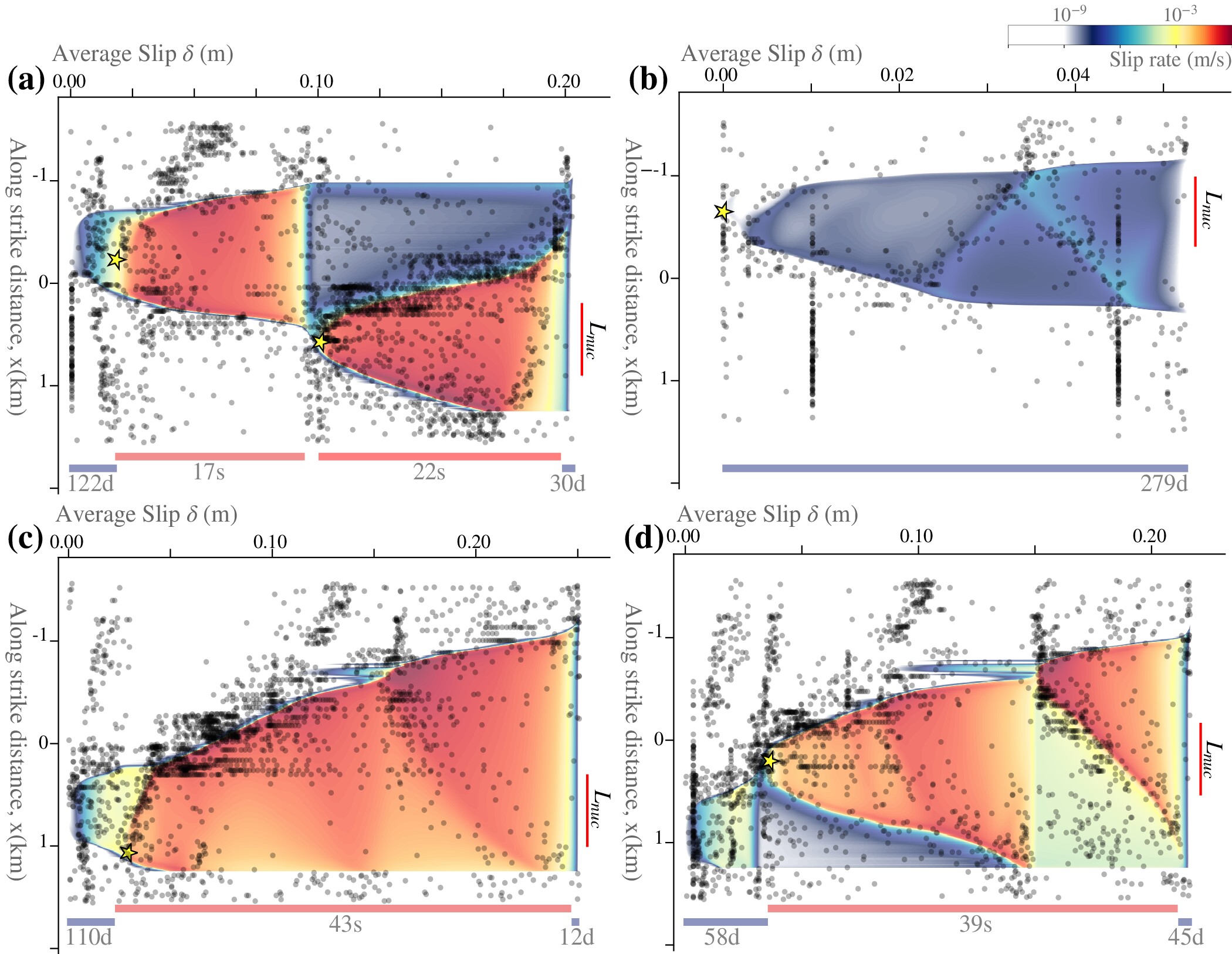}
	\caption{Sequences of slip rate profiles on the main fault. Top x-axis is average slip
		accumulated during the specific time sequence. Left y-axis represents the along-strike
		distance on the main fault. Horizontal colored lines show the specific durations for
		different phases of the rupture sequence, with blue lines indicating slow phases and red
		lines indicating fast phases. Nucleation and afterslip phases are indicated by two blue
		lines at the beginning and end of fast events. The colormap represents slip rate. Black
		circles represent events detected on the fractures, and projected onto the main fault
		strike. Yellow stars represent the epicenters of the different events. Panel a) shows
		complex partial ruptures on the main fault. Panel b) shows one slow slip event. Panel c)
		shows a full rupture on the main fault. Panel d) shows an event, nucleated from one end of
		the fault and accelerated in a cascading process.}
	\label{slip-rate-profile}
\end{figure}

Having explored temporal dynamics, we now examine spatial complexity of the seismic events recorded
on the main fault (Figure~\ref{slip-rate-profile}). We delve into the spatial complexity of seismic
events recorded on the main fault by looking at the slip rate profiles along strike as a function of
spatially averaged cumulative slip during the event. In addition, we project all off-fault activity during this event
onto the main fault and denote them by simple black circles. For our analysis, we carefully selected
several representative events from the case study, encompassing a diverse range of slip occurrences.
This includes partial ruptures, full ruptures, and slow slip events, each characterized by distinct
magnitudes and durations. For the fast ruptures, we present both the nucleation phase, where the
slip rate starts to accelerate from an arbitrarily small threshold, i.e., $10^{-9}$ m/s, prior to
reaching some geodetically detectable threshold, i.e., $10^{-6}$ m/s, and the afterslip phase, where
the slip rate decelerates to the same limits. Figure~\ref{slip-rate-profile}a features two partial
ruptures, each encompassing around half of the main fault. Notably, the second rupture (moment
magnitude $M_w = 4.70$) nucleates at the edge of the first one ($M_w = 4.56$), demonstrating a
cascading effect. Despite the rapid rupture of coseismic phases, which last 17 and 22 seconds,
respectively, the nucleation phase lasting 122 days and the afterslip phase taking place over 30
days are significantly slower.

In Figure~\ref{slip-rate-profile}b, one slow slip event is showcased, rupturing only part of the
main fault. This slow slip event spans 279 days with a moment magnitude $M_w = 4.48$. It is
essential to acknowledge the subjectivity in detecting slow events due to the slip rate threshold
used (here defined by slip rates between $10^{-9}$ and $10^{-3}$ m/s). This threshold effect will be
discussed later in the text. We observe continuous seismic activity on the off-fault fractures
accompanying the slow event on the main fault, as previously seen in Figure~\ref{moment-rate}c.
Figure~\ref{slip-rate-profile}c unfolds a full rupture on the main fault with a moment magnitude of
$M_w = 5.03$, revealing nuanced variations in slip rate acceleration and deceleration within the
rupture and along the fault's strike. The nucleation and afterslip phases of this event last 110 and
12 days, respectively. Figure~\ref{slip-rate-profile}d features another full rupture on the main
fault ($M_w = 4.95$), preceded by a nucleation phase that ruptures around one-third of the main
fault over 58 days. The rupture accelerates from the left corner of the nucleation patch. Then, the
full rupture on the main fault lasts for 39 seconds; after a 45-day afterslip phase, the slip rate
falls below $10^{-9}$ m/s. The figure illustrates a distinct cluster of off-fault events following
the rupture front on the main fault.

Beyond demonstrating that slow and fast events emerge naturally within the same fault system under
identical constitutive parameters, our model is also able to produce spatially localized regions
where slow slip is persistent. This emerges naturally from the local geometric configuration of the
main fault and surrounding off-fault fractures, without requiring imposed frictional or hydraulic
heterogeneity. The coexistence of slow and fast slip events on the same fault or within the same
region (see Figure \ref{segregation}), as produced by our simulations, is also supported by a
growing body of observational, geological, and laboratory evidence. Seismological observations
document the coexistence or close spatial association of slow slip, tremor, very-low-frequency
earthquakes, and regular earthquakes in several tectonic settings \parencite[e.g.][]{Kato2012,
Johnson2012, Ito2013, Ruiz2014, Tsang2015, Lin2020, Villafuerte2025, Woods2024}. Geological studies
indicate that deformation mechanisms associated with slow and fast slip can operate within the same
fault zones over geological timescales \parencite[e.g.][]{Fagereng2019}, and laboratory experiments
demonstrate the coexistence and interaction of slow and fast slip under controlled conditions
\parencite[e.g.][]{Aben2023}.

\subsection{Slow to fast events across scales}

\begin{figure}[!ht]
	\centering
	\includegraphics[width=\textwidth]{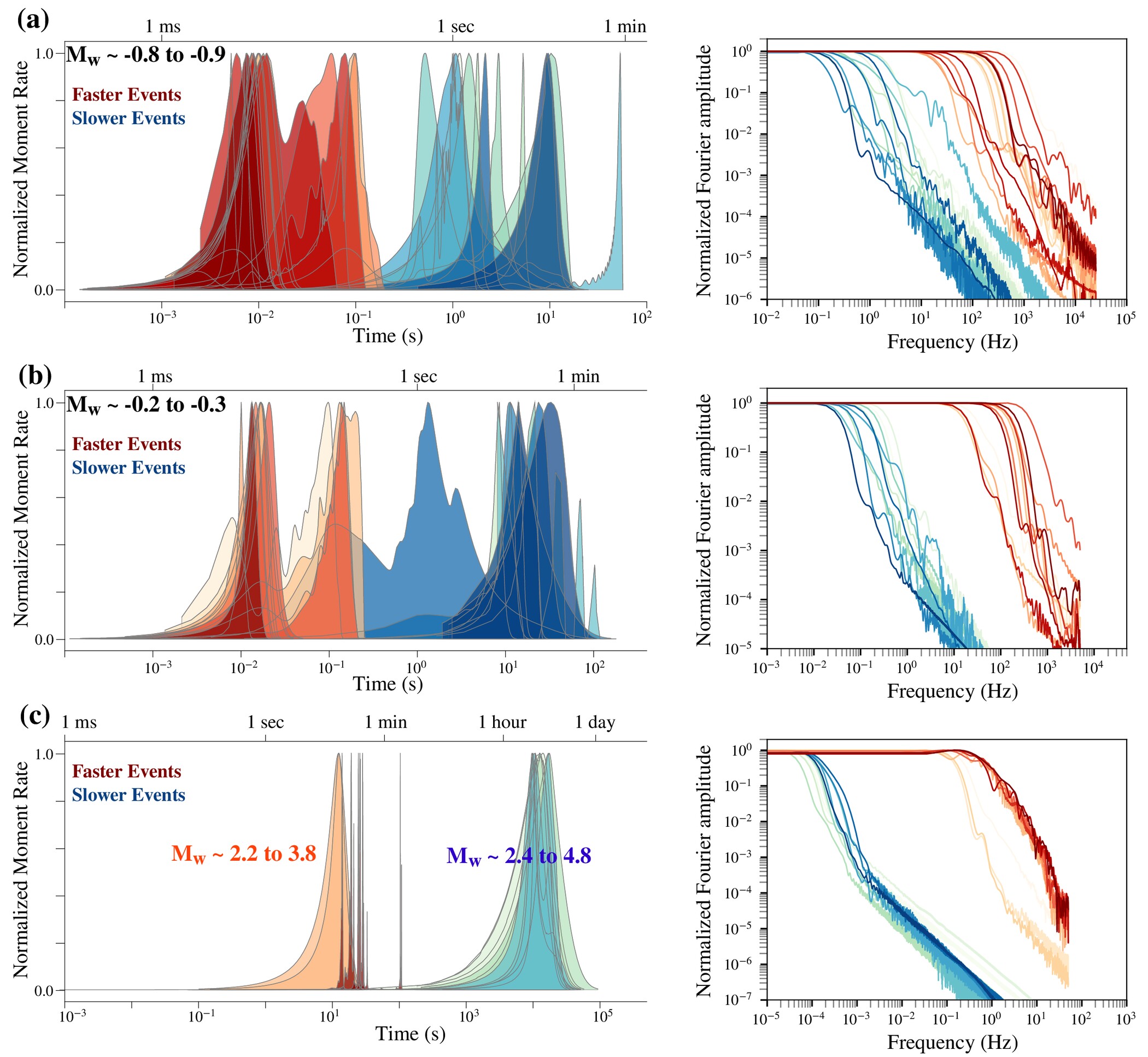}
	\caption{Moment-rate functions (MRFs) (left panels), and their spectra (right panels), across
	different $M_w$ ranges for both off-fault and main fault events. (a) Selected MRFs from slow and
	fast events in the off-fault region within a narrow range of $-0.9 \leq M_w \leq -0.8$. (b) Same
	as (a), but for $-0.3 \leq M_w \leq -0.2$. (c) MRFs from main fault events, and their spectra,
	with fast ruptures in the range $2.2 \leq M_w \leq 3.8$ and slow slip events in the range $2.4
	\leq M_w \leq 4.8$.}
	\label{MRF_spectra}
\end{figure}

We present a general overview of the source time functions (STFs) generated within our fault system
model, focusing on their variability in moment magnitude, duration, and spectral characteristics.
The analysis includes moment-rate functions (MRFs) from both slow and fast events, considering a
representative fault volume geometry in which off-fault fractures are oriented parallel to the main
fault (unlike the case discussed earlier where the off-faults are optimally oriented right-lateral
faults). 

While we get a continuum of slow and fast events for all off-fault geometries, we selected this
particular geometry because it exhibits a remarkably broad and diverse range of temporal behaviors,
capturing both small-magnitude slow events and large slow slip events occurring on both off-faults
and the main fault. This suggests that adjusting frictional parameters could potentially produce
similar behavior across different geometries, though this requires further investigation.

\begin{figure}[!ht]
	\centering
	\includegraphics[width=0.75\textwidth]{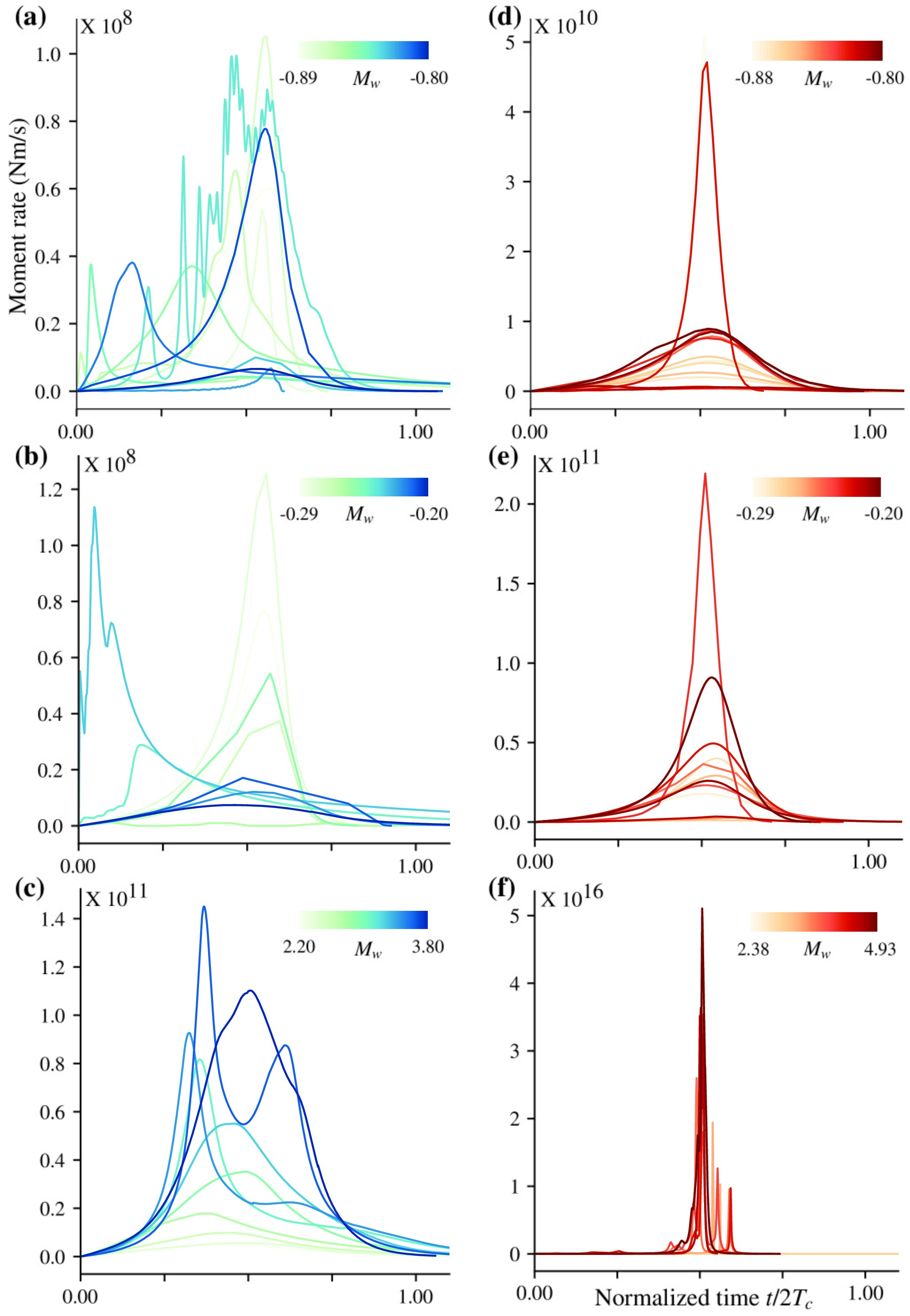}
	\caption{MRF shapes for slow and fast events after normalization by centroid time  with colors
	indicating moment magnitude. Panels (a) and (b) correspond to slow events in the off-fault
	region for the same magnitude ranges as in Figures~\ref{MRF_spectra}a and \ref{MRF_spectra}b.
	(c) shows slow events on the main fault. Panels (d) and (e) correspond to fast off-fault events
	with the same $M_w$ ranges as in (a) and (b). (f) Fast events on the main fault.}
	\label{MRF_asymm}
\end{figure}

Figures~\ref{MRF_spectra}a and \ref{MRF_spectra}b show selected MRFs from off-fault events within
narrow magnitude ranges: $-0.9 \leq M_w \leq -0.8$ and $-0.3 \leq M_w \leq -0.2$, respectively.
These MRFs exhibit a continuous and broad distribution of slip durations, with significant
variability in their shape, ranging from abrupt moment release to smoother and more gradual
evolutions. While the fast events are classical earthquakes, the slower events can easily be called
LFE's and VLFEs. To confirm this, we need to radiate these events dynamically and look at their
far-field waveforms, which we plan to do in future work. Although not explored here, our model
framework could also explain tectonic tremors as successive ruptures of shear cracks as recently
proposed by \textcite{Yabe2025}. This diversity highlights the influence of structural complexity
and distributed deformation in controlling rupture behavior, emphasizing that seismic complexity
arises not only from the main fault, as shown previously, but also from off-fault activity.

Figure~\ref{MRF_spectra}c shows MRFs of events on the main fault, which exhibit a bimodal behavior.
Due to the greater extent and roughness, the main fault hosts the largest magnitude events,
characterized by more complex STFs, including multiple acceleration and deceleration phases during
the ruptures. The magnitude range considered here includes both large-magnitude, fast ruptures,
partial or full, ($2.2 \leq M_w \leq 3.8$) and long-duration slow slip events ($2.4 \leq M_w \leq
4.8$).

We now examine the spectral characteristics of these events. The corresponding source spectra for
the same events show, as expected, that the longer durations of the slow events lead to
significantly lower corner frequencies than fast events across all three magnitude ranges, on both
the main and off-faults. Recently, \textcite{Wang2023} analyzed the S-wave displacement spectral
signature of low-frequency earthquakes in the Nankai Trough, after correcting for empirically
derived attenuation, and concluded that the spectra are consistent with the classical earthquake
model with vastly different rupture velocities and stress drops. Our results align very well with
the conclusions drawn by \textcite{Wang2023}. Beyond the corner frequency, the spectral decay of
slow events appears similar to that of fast events. However, the high-frequency content is less
reliably estimated, as it would be predominantly influenced by the fully dynamic response of both
event types, an aspect that our quasi-dynamic model does not account for, but which we aim to
address in future investigations. To further examine MRF characteristics across these events, we
normalized the time axis of each MRF with respect to its centroid time, $T_c$
(Figure~\ref{MRF_asymm}). The centroid time is the moment rate weighted average time
\parencite{Dziewonski1981, Duputel2013} and is defined as $T_c = {\int_{t_1}^{t_2} t \dot{M}_0(t) \,
dt}/{\int_{t_1}^{t_2}\dot{M}_0(t)}$. Here $\dot{M}_0(t)$ is the moment rate, and $t_1$ and $t_2$
correspond to the onset and termination times of the MRF, defined by when the slip rate crosses a
detection threshold.

One key distinction between slow and fast STFs lies in their symmetry. MRFs of slow events
(Figures~\ref{MRF_asymm}a–c) tend to be more asymmetric, typically exhibiting a sharp rise followed
by a longer, gradual decay—indicating a positively skewed moment release. In contrast, fast
earthquakes generally show more symmetric MRFs with a single, well-defined peak
(Figures~\ref{MRF_asymm}d–f). However, for the largest slow slip events (Figure~\ref{MRF_asymm}c),
the STFs of both fast and slow events become more irregular, and the asymmetry for some events is
less distinguishable. Consistent with the findings of \textcite{Meier2017}, our results suggest that
fast earthquakes tend to have simpler MRFs, with fewer subevents and a well-defined peak and gradual
decay (Figures~\ref{MRF_asymm}d–f). While this section focuses on a specific fault geometry, similar
results were obtained for all the 40 fault volume configurations tested in this study.

\section{Statistical and Scaling Laws of the Events in the Catalog}

In this section, we demonstrate the ability of our numerical model to replicate observed empirical
statistical relationships and scaling laws governing seismic activity. These empirical relationships
include the Omori law for the decay in the rate of aftershocks, Gutenberg-Richter
magnitude-frequency distribution, the inverse-Omori law for foreshock escalation, moment-area
scaling, and moment-duration scaling. We posit that any reasonable `digital twin' of a fault system
should be able to reproduce these empirical laws in addition to the range of slip dynamics presented
earlier. Through this analysis, we plan to show that our fault volume model not only produces a
diverse range of slip dynamics but also satisfies key empirical laws observed in natural seismicity.
We show that for each geometrical configuration and friction parameters, the above listed empirical
laws are satisfied (see Figures \ref{moment-rate}e, \ref{stats},
\ref{scalingMT}). For the sake of brevity here we show the robustness of these
results across all the different cases of off-fault orientations and frictional parameters. 

\subsection{Omori Law}\label{sec:omori} 

\begin{figure}[t!]
	\centering
	\includegraphics[width=\textwidth]{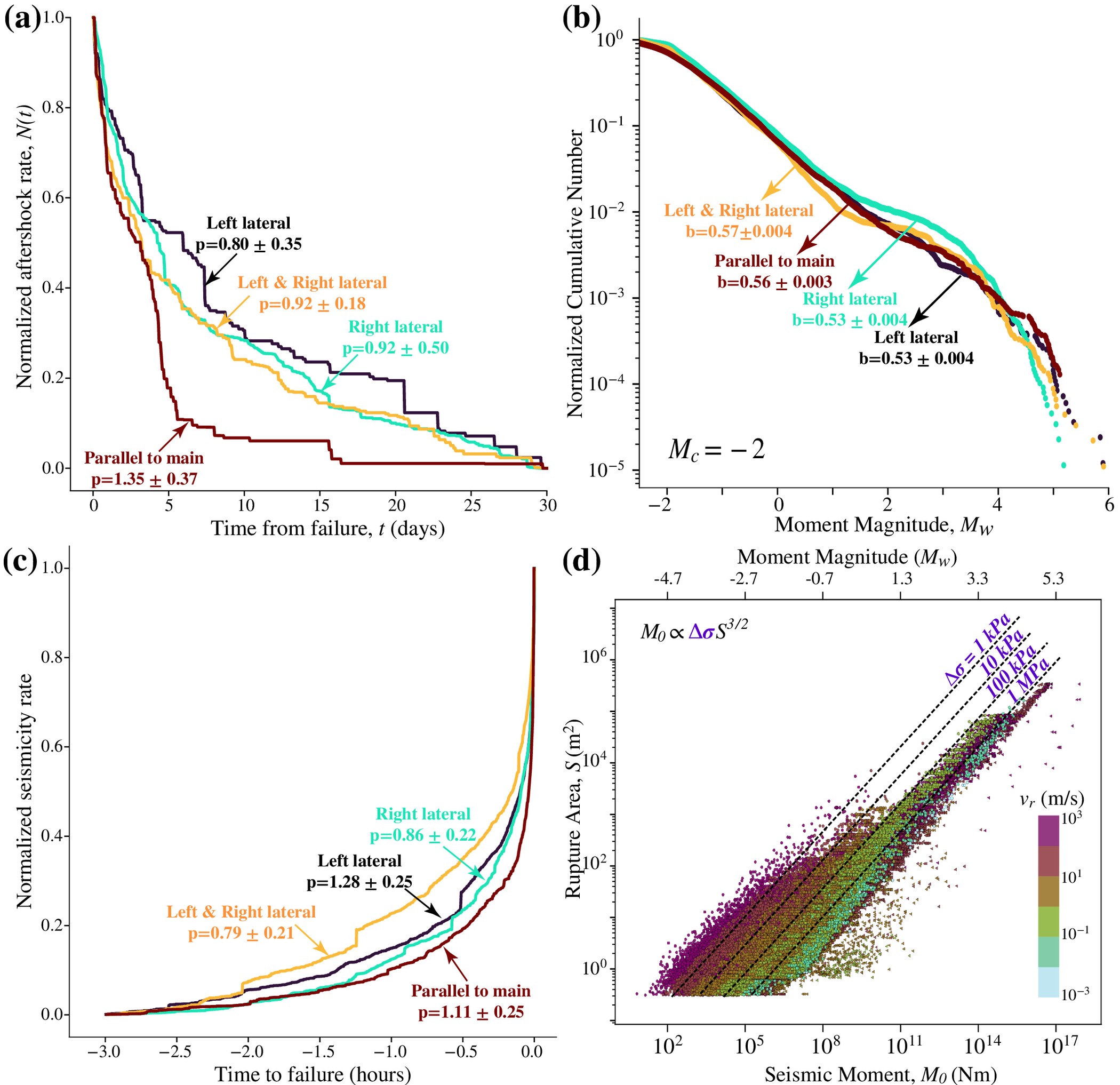}
	\caption{(a) Omori Law: Effect of off-fault fracture orientation. The average Omori decay curve
	over a period of 30 days is calculated from the stacked curves of different fracture
	orientation, and the average $p$ coefficient is reported (b) Magnitude Distribution: Effect of
	off-fault fracture orientation on the b-value found from stacked catalog of fast ruptures with
	same off-fault fracture orientation (c) Inverse Omori: Effect of off-fault fracture orientation
	(d) Moment-area scaling: The $a/b$ ratio of the off-fault fractures is 0.5.}
	\label{globalstats}
\end{figure}

In seismological studies, the Omori law characterizes the temporal decay of aftershock activity
following a mainshock event. Specifically, it states that the number of aftershocks decreases
inversely with time after the mainshock, following a power-law decay as follows: $n(t) = k/(c+t)^p$
where $k$ is aftershock productivity, $c$ is a time offset, and $p$ is a constant that describes the
decay rate and typically falls in the range 0.8–1.2 in most cases \parencite{omori1894, Utsu1995}.
To evaluate the Omori law in our simulations, for each orientation distribution of the off-fault
fractures, we stack the aftershock sequences following each major rupture on the main fault. We then
fit the stacked sequences to the Omori law to estimate the $p$ exponent. Figure~\ref{globalstats}a
illustrates the Omori law coefficient $p$ across different fracture orientations when $a/b = 0.5$.
The coefficient remains consistently around 1 across all orientations, indicating a uniform behavior
regardless of these parameters. While off-fault orientation does influence the duration of the
aftershock period (see the case when off-faults are aligned with the main fault), we defer this
investigation to future studies as it falls outside the scope of our current research.

\subsection{Gutenberg-Richter Magnitude Frequency Distribution}
\label{scalingsection}

Figure~\ref{globalstats}b illustrates the magnitude-frequency distribution of ruptures across
different fracture orientations. The x-axis denotes the magnitude range, while the y-axis represents
the frequency of occurrences. The distribution follows the Gutenberg-Richter law, exhibiting a
logarithmic relationship between magnitude and frequency. This distribution provides crucial
insights into the relative occurrence rates of seismic events of varying magnitudes, offering
valuable information for seismic hazard assessment and earthquake forecasting. To estimate the
b-value, we utilize the maximum likelihood method \parencite{Aki1965}, compensating for the binning
error \parencite{Marzocchi2009}. We note that $M_w = -2$ approximately serves as the completeness
magnitude of our catalog. All plotted distributions conform to the Gutenberg-Richter law, displaying
a similar $b$-value across orientations. Interestingly, there is no apparent influence of off-fault
fracture orientation on the magnitude-frequency distribution, suggesting that this aspect does not
significantly impact slip dynamics.

As shown by \textcite{aki1981}, a b-value between 0.5 and 1 can be envisaged when the fractal
dimension is between 1 and 2 where the fault lines are distributed along a plane. This needs to be
examined in much more detail. Although there is still room to investigate the convexity of the
magnitude–frequency distribution using truncated G–R distributions \parencite{Huang2015}, we leave
this investigation for future studies. It is important to note that our primary objective is not to
study the origin of the b-value derived from our fault volume model. Rather, our focus is to
demonstrate that considering the dynamics of a fault volume with fractures surrounding the main
fault results in a logarithmic relationship between magnitude and frequency, consistent with the
observed Gutenberg-Richter law in nature. While we suspect that the b-value may be influenced by
factors such as the power-law distribution of fracture length and density distribution of fractures
along fault-normal distance, we defer this exploration to future studies, as it falls beyond the
scope of our current investigation.

\subsection{Inverse Omori Law}\label{sec:invomori}

In addition to the Omori-law-like reduction in the earthquake rate after a given event
(section~\ref{sec:omori}), the catalog of events in this paper's simulation follows the empirical
inverse Omori law that characterizes the increase in foreshock activity preceding a major rupture
event. Unlike the aftershock decay described by the traditional Omori law, the inverse Omori law
predicts a gradual rise in foreshock activity leading up to a significant seismic event, where
foreshock rate increases as an inverse of the time to the mainshock as follows: $n(t) = 1/(c +
\Delta t)^p$ \parencite{Jones1979, Shearer2023}, where $p$ is a constant that describes the increase
rate of foreshocks and is typically around 1. In Figure~\ref{globalstats}c, we show the increasing
rate of seismicity in a 3-hour window prior to the main shock, grouped by fracture orientation. The
increasing rate of seismic events off-fault exhibits an inverse Omori law, with $p$-values close to
1. The alignment with the Omori law underscores its applicability in describing the temporal
evolution of aftershock activity within our simulation. Similarly, the consistent trend of foreshock
rate increase observed across our simulated seismic events aligns well with the predictive power of
the inverse Omori law, demonstrating its ability to capture precursory behavior preceding major
seismic events and laboratory earthquakes as shown in \textcite{Marty2023}.

\subsection{Moment-Area Scaling Law}

Next, in Figure~\ref{globalstats}d, we explore the relationship between seismic moment ($M_0$) and
rupture area ($S$) for both fast and slow ruptures. Notably, while determining rupture area ($S$) in
nature can often be challenging, our model offers a straightforward approach as we have precise
knowledge of fault behavior. Given our 2D model, we assume a circular rupture shape, simplifying the
calculation of rupture area as equal to the square of rupture length. The figure reveals a scaling
relation of $M_0 \propto S^{3/2}$, mirroring patterns observed in natural seismicity
\parencite{Kanamori2001}. It is essential to note that this scaling law is consistent across both
fast and slow ruptures. This scaling law underscores the fault volume model's fidelity in capturing
the spatial distribution of seismic energy release across various rupture sizes. Importantly, the
observation that both fast and slow ruptures exhibit almost the same scaling between moment and
rupture area is noteworthy. Given that seismic moment relates to stress drop and rupture area as
$M_0 = C \Delta \sigma S^{3/2}$ \parencite{eshelby1957,Kanamori2001,Madariaga2009}, where $C$ is a
geometric constant ($C = 16/7\pi^{3/2}$ for a circular crack), the variability in stress drop must
be smaller than that of rupture velocity. In fact the above relationship can be rearranged to show
that $\overline{V} \mu /\overline{v_r}\Delta\sigma$ is a constant. Here $\overline{V}$ is the
average slip rate of the event and $\overline{v_r}$ is the average rupture velocity of the event. In
our simulations, average sliprates vary over 8 to 9 orders of magnitude between slow and fast
events, while rupture velocities vary over 5 to 6 orders of magnitude. This implies that stress
drops should vary over 2 to 3 orders of magnitude across the entire catalog of events. It is also
clear that average slip rate, stress drop and average rupture velocity are three fundamental
quantities of an event that cannot vary independently and are linked through the above relationship.
\textcite{Supino2020}, for instance, found that low-frequency earthquakes propagate more slowly and 
have lower stress drops than regular earthquakes.

\subsection{Moment-Duration Scaling Law}

\begin{figure}
	\centering
	\includegraphics[width=\textwidth]{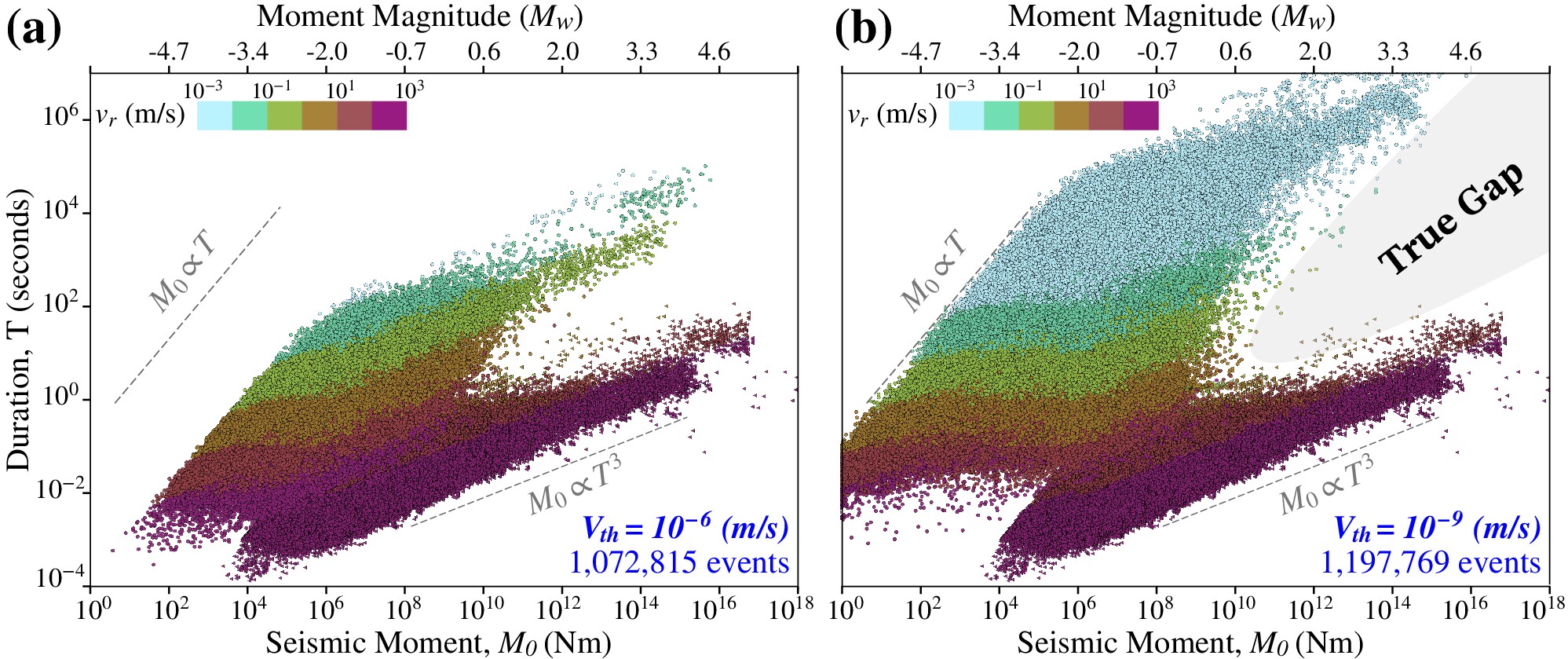}
	\caption{Moment-duration scaling for the comprehensive catalog compiled from all simulations :
	a) with a threshold of detection of slow ruptures at $10^{-6}$ m/s, showing a $M_0 \propto T^3$
	relation for fast and slow ruptures; b) with a threshold of detection of slow ruptures at
	$10^{-9}$ m/s. Events are color-coded based on rupture velocity $v_r$.}
	\label{full-stat-MT}
\end{figure}

In Figure~\ref{full-stat-MT}, we delve into the moment-duration scaling, a crucial aspect of slow
and fast ruptures. Observations in natural seismicity have highlighted a cubic relation between
moment and duration for fast ruptures: $M_0 \propto T^{3}$ \parencite{Kanamori1975, Kanamori2004,
Houston2001}. However, for slow ruptures, the situation appears more complicated, with some
observations suggesting a linear relation: $M_0 \propto T$ \parencite{Ide2007, Ide2023}, while
others indicating a cubic one \parencite{Gomberg2016, Michel2019}. To investigate this further, we
assess whether the detection threshold for slow ruptures influences the observed scaling
relationship. Specifically, we examine Figure~\ref{full-stat-MT}b, where the detection threshold is
set to $10^{-9}$ m/s, and compare it to Figure~\ref{full-stat-MT}a, which uses a threshold of
$10^{-6}$ m/s. This comparison allows us to evaluate how the choice of detection threshold affects
the apparent moment–duration scaling. Notably, the latter threshold is still three orders of
magnitude smaller than the threshold for fast ruptures, set at $10^{-3}$ m/s. The color scale
represents the rupture velocity. The bottom x-axis displays seismic moment ($M_0$), while the top
x-axis shows moment magnitude ($M_w$) for clarity and comparison with observational data. For the
fast ruptures, we observe a clear cubic scaling ($M_0 \propto T^3$) between moment and duration,
consistent with previous findings. 

\begin{figure}[!ht]
	\centering
	\includegraphics[width=0.75\textwidth]{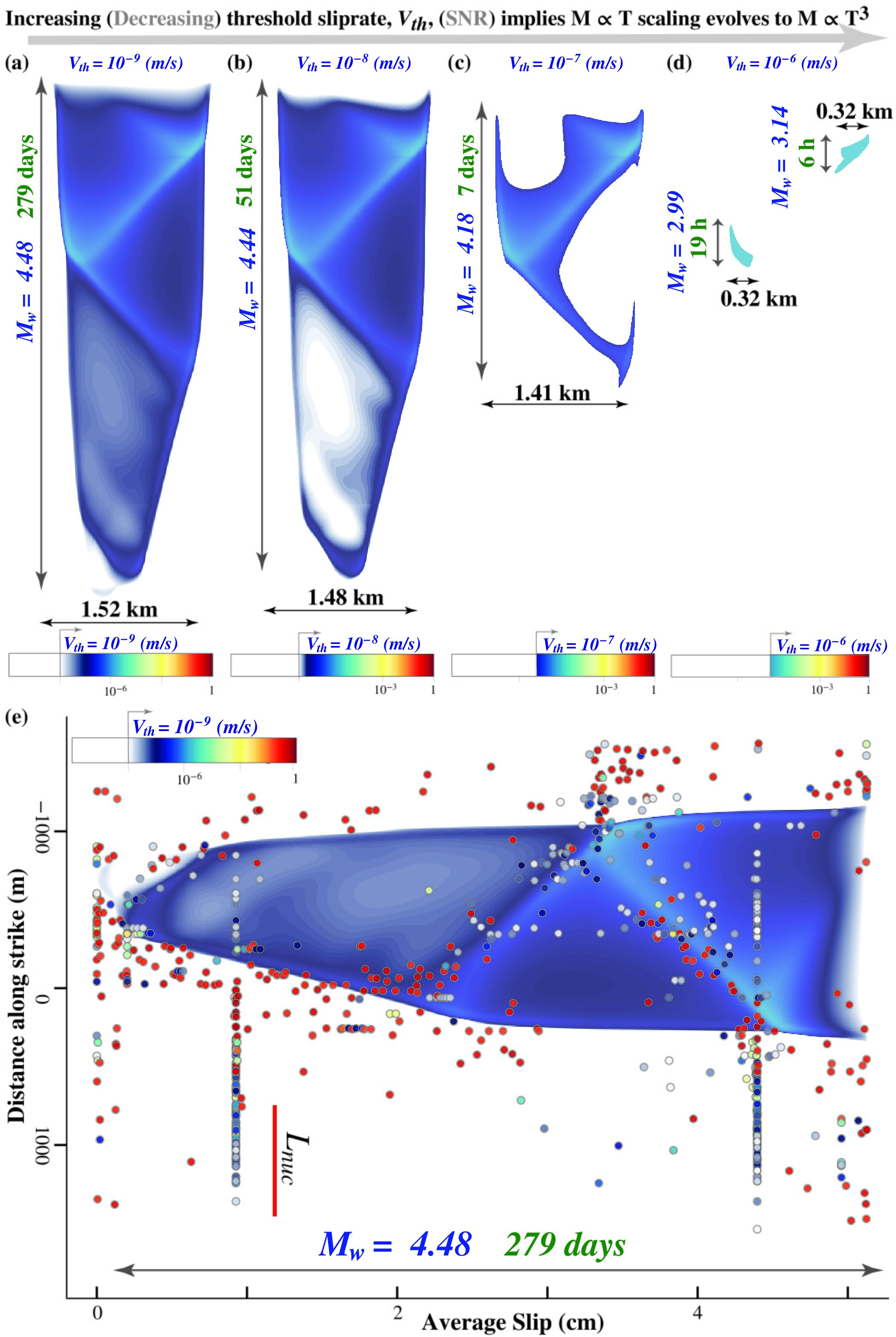}
	\caption{Effect of slip rate threshold ($V_{th}$), or signal to noise ratio (SNR), on the
	inference of the magnitude and duration of a slow slip event. The slip rate is plotted with four
	different minimum detection levels: (a) \(10^{-9}\) m/s, (b) \(10^{-8}\) m/s, (c) \(10^{-7}\)
	m/s, and (d) \(10^{-6}\) m/s. (e) Same as (a) with off-fault seismicity. Circles represent
	ruptures on the off-fault fractures, and are color-coded with respect to their maximum slip
	rate.}
	\label{threshold}
\end{figure}

However, the behavior of slow ruptures varies between the two detection thresholds. With a lower
threshold (Figure~\ref{full-stat-MT}b), the durations of slow events are larger and we also obtain
more events compared to the higher threshold case (Figure~\ref{full-stat-MT}a). The overall trend of
the slow events thus changes as a function of the detection threshold. This discrepancy suggests
that the scaling between moment and duration for slow ruptures is highly dependent on the detection
threshold in our numerical model, akin to the sensitivity or threshold settings of field recording
instruments or detection strategies employed on field data. In a recent work
\textcite{Costantino2026}, similar threshold-dependent effects were observed, in natural settings,
when applying deep learning denoising techniques to reveal a continuum of slow slip events in the
Cascadia subduction zone, demonstrating that detection threshold fundamentally shapes our
understanding of slip behavior. The effect of the detection threshold, ranging from $10^{-9}$ m/s to
$10^{-6}$ m/s, on the detectable moment and magnitude of an example slow-slip rupture is shown in
Figure~\ref{threshold}a-d, highlighting how inferences drawn from the duration and moment of a
slow-slip rupture could be changed if the lowest slip rate at which the rupture is initiated cannot
be detected. 

While the slip rate on the main fault may be very low and remain undetectable, the pattern of
off-fault seismicity migration provides valuable information about the rupture process on the main
fault, Figure~\ref{threshold}e. The dot symbols on the plot represent off-fault events projected
onto the main fault, marking the extents of the slow slip rupture. Further analysis is warranted to
assess how reliably the pattern of off-fault seismic activity can be used to infer the magnitude and
duration of a slow event, and we defer this investigation to future studies.

Since each of the events is color-coded based on its average rupture velocity in
Figure~\ref{full-stat-MT}, it is apparent that slow or fast events are identical in nature except
for their rupture velocity. In fact, it is evident that $M_0 \propto T^3$ and $M_0 \propto T$ are
limiting bounds in the moment-duration scaling as proposed by \textcite{Ide2023}. This suggests that
slow ruptures are simply fast ruptures with different slip rates, rupture velocities, and stress
drops. This also implies that the slow ruptures do not emerge due to a special frictional
constitutive law since it is the same mechanism that generates fast ruptures as well. What thus
produces the difference in rupture velocity is the spatiotemporally complex traction on the main
fault due to the presence of the fault volume. We also observe small events tend to have a continuum
of rupture velocities; however, as events become larger, a distinct gap emerges between fast and
slow ruptures. This gap mirrors observations reported by \textcite{Ide2023}, suggesting events that
are challenging to detect or exceedingly rare. Our numerical model's ability to detect these events
with confidence reinforces the presence of this gap, raising the possibility of a mechanical
constraint where large events predominantly exhibit either fast or slow rupture velocities, rather
than a continuum. This strongly suggests that both fast and slow ruptures do originate from the
same mechanical model, despite their differing rupture velocities. 

\subsection{Localization and Migration of Seismicity}

In this section, we address the question of localization and delocalization of deformation
throughout the seismic cycle, aiming to understand how deformation occurs in a fault volume over a
seismic cycle. Over decadal timescales during the interseismic period, elevated background seismic
activity may reflect progressive damage accumulation across a broad region that will eventually host
a major earthquake. In several large events including the 1992 $M_w$ 7.3 Landers, the 1999 $M_w$ 7.1
Hector Mine, the 2019 $M_w$ 7.1 Ridgecrest, and the 2023 $M_w$ 7.8 Kahramanmaraş earthquakes,
seismicity has been observed to migrate and concentrate toward the eventual rupture plane over a few
years to several months preceding the mainshock. This localization appears to facilitate the
interaction and coalescence of fault segments and fractures that delineate the future principal slip
zone \parencite{BenZion2019, BenZion2020, Pritchard2020, Kato2020, Kwiatek2023, NunezJara2025}.

\begin{figure}
	\includegraphics[width=\textwidth]{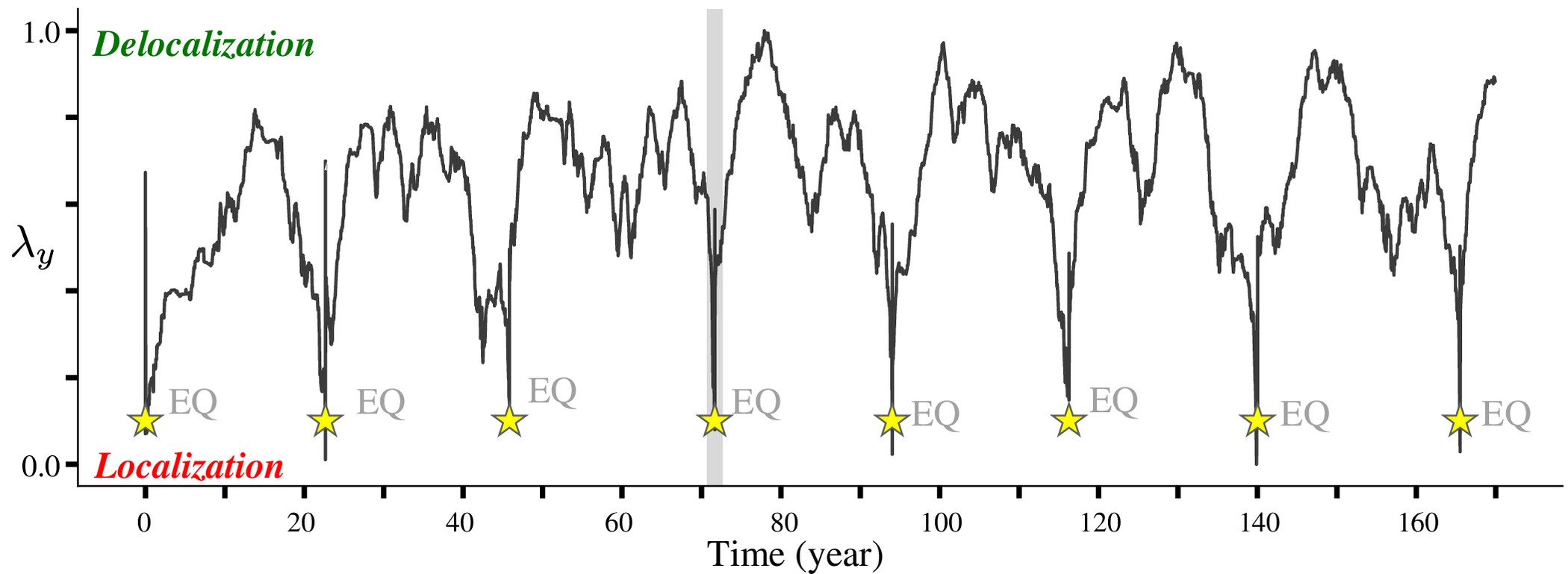}
	\caption{Localization and delocalization of off-fault seismic activity, measured by the standard
	deviation of hypocentral distances from the main fault in sequential 500-event batches,
	$\lambda_y$. The timing of earthquakes is indicated by yellow stars. The standard deviation
	values are normalized by their maximum value over all batches. The gray region highlights the
	time window shown in Figure~\ref{migration}.}
	\label{localization}
\end{figure}

In our simulations, we quantify this localization by looking at the location of off-fault events
over time. To this end, we calculate the standard deviation of hypocentral distances from the main
fault ($\lambda_y$) in sequential 500-event batches and normalize the values between 0 and 1. The
events tend to localize toward the main fault plane when $\lambda_y \to 0$ and delocalize into the
damage zone when $\lambda_y \to 1$. Figure~\ref{localization} shows the time series of $\lambda_y$.
A reduction in this time series before the main fault earthquake, followed by an increase after it,
is clearly distinguishable, highlighting the localization of seismicity onto the main fault as the
time of the mainshock approaches, and its delocalization after the occurrence of the earthquake.

\begin{figure}[!ht]
	\includegraphics[width=\textwidth]{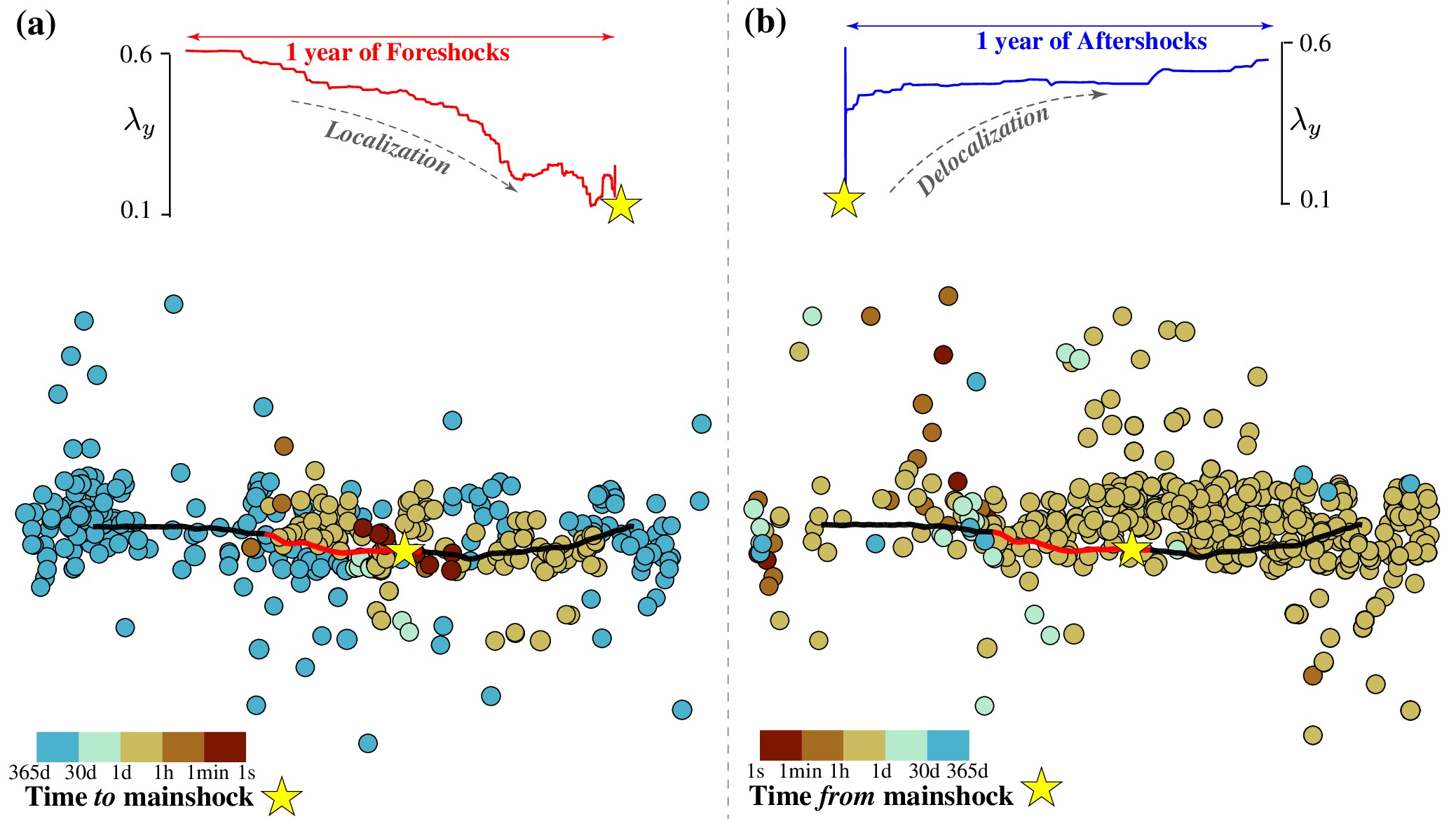}
	\caption{ Zoomed-in view of the gray-hatched window shown in Figure~\ref{localization},
	illustrating the standard deviation of the fault-normal component of earthquake locations over a
	one-year window before (a) and after the earthquake (b)) : Migration of seismicity toward the
	future epicenter during the foreshock period, followed by a return to background seismicity
	after the earthquake. Events are shown as circles, color-coded by time before and after the
	rupture.}
	\label{migration}
\end{figure}

A closer look at the location of off-fault events reveals a migration toward the epicenter of the
upcoming mainshock. We observe this feature somewhat systematically in our simulations but show only
the visually clearest example below. Figure~\ref{migration} shows the location of off-fault events
during the time interval highlighted by the gray hatch in Figure~\ref{localization}. Specifically,
Figure~\ref{migration}a shows a one-year window of foreshocks, while Figure~\ref{migration}b shows a
one-year window of aftershocks. Events are shown by circles, color-coded based on time to failure in
panel (a) and time from failure in panel (b). This figure clearly illustrates the migration of
off-fault events toward the epicenter, emphasizing the importance of refined event catalogs for
detecting such patterns and their potential forecasting value as a mainshock approaches. Following
the rupture on the main fault, $\lambda_y$ increases, and off-fault seismicity no longer exhibits a
migration pattern.

\subsection{Apparent Diffusion}

\begin{figure}[!ht]
	\centering
	\includegraphics[width=\textwidth]{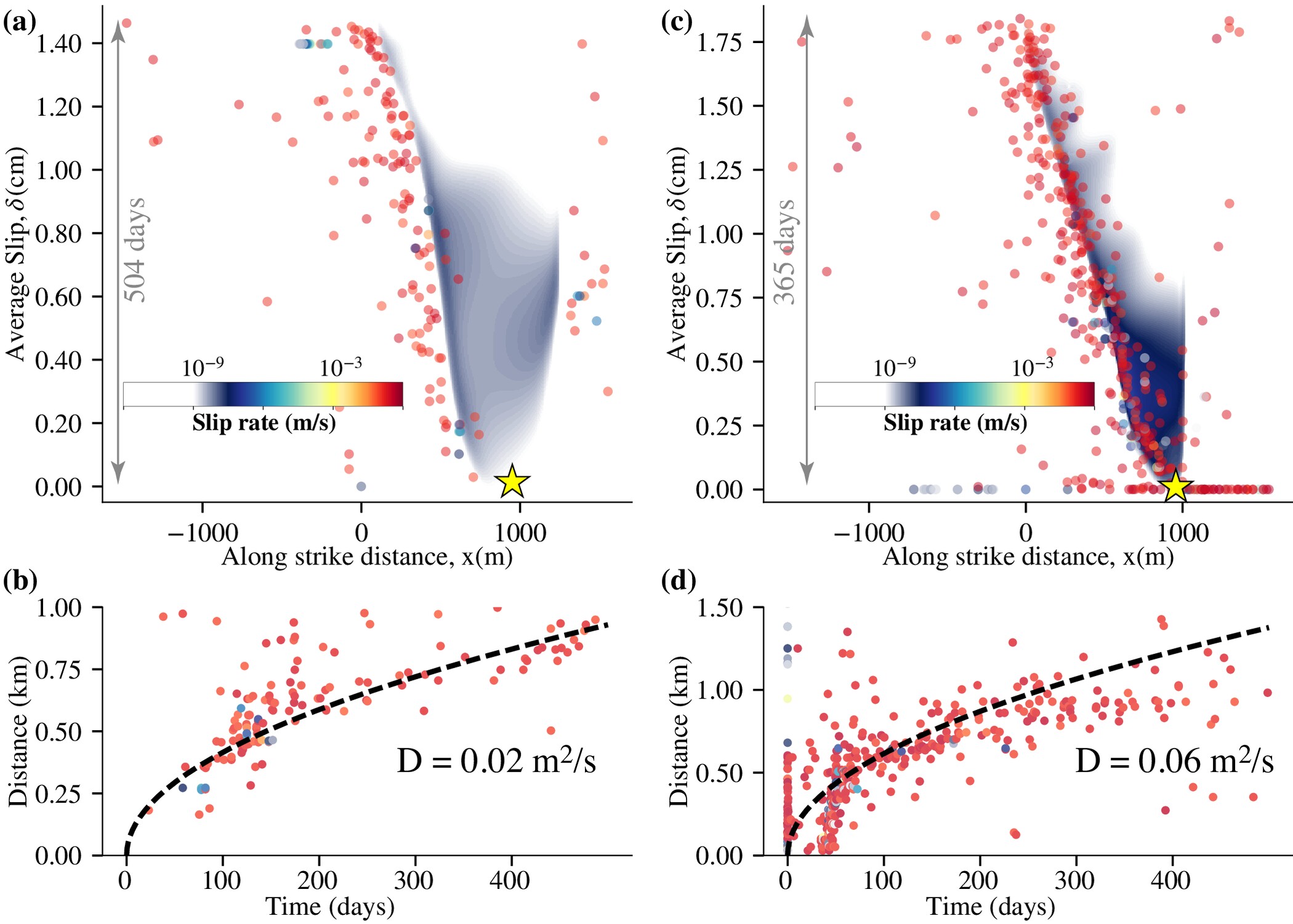}
	\caption{Off-fault Seismicity during two examples of slow-slip ruptures. a-c) Slip rate profiles
    of two slow slip events. b-d) The temporal migration of off-fault seismicity follows a distance
    from epicenter ($r$)–time ($t$) evolution (black dashed curve) similar to that of a diffusion
    front ($r \propto D\sqrt{t}$). The relative distance is calculated as the distance between the
    off-fault event and the detected nucleation point on the main fault, using a slip rate threshold
    of $10^{-9}$ m/s. The corresponding diffusion coefficient ($D$) is noted at the top of each
    curve.}
	\label{diffusion}
\end{figure}

It has been observed that the migration of seismicity in some fault zones follows a specific
space-time evolution, similar to the one obtained from fluid diffusion \parencite[e.g.,][]{Danr2024}. In
other words, the distance $r$ from the epicenter versus the time from the main shock $t$ is observed
to be $r \propto D\sqrt{t}$. In our simulations, we observe the same evolution despite the absence of any underlying fluid
diffusion process. This observation is illustrated in Figure~\ref{diffusion}, where two slow-slip
ruptures are shown (panels a,c) along with seismic events occurring off-fault overlaid on the slip
rate panels. The relative distance is calculated as the distance between the off-fault event and the
detected nucleation point on the main fault, using a slip rate threshold of $10^{-9}$ m/s. The
relative time is determined by the time difference between the occurrence of the off-fault event and
that of the main fault. This off-fault seismicity migration is shown in panels (b,d). The values of
the inferred diffusivity parameter $D$ (0.02–0.06 m$^2$/s) fall within the observed range for the
fluid diffusivity parameter \parencite{Amezawa2021}.

Our simulations reveal an apparent diffusive process driven by slow slip rather than fluids,
demonstrating that seismicity migration following $r \propto \sqrt{t}$ does not necessarily imply
underlying fluid-driven processes. This behavior can also emerge from the physics of interacting
faults governed by rate-and-state friction along with a slow slip event occurring on the main fault.

\section{Discussion}
These findings, when taken in total, highlight the fault volume framework's capacity to capture
global, fault-related scaling laws and statistics, both on the main fault and within the fractured
medium. Moreover, the fault volume model accurately reproduces diverse slip dynamics, slow to fast
earthquakes across scales, all within a unified framework. Our simulations also replicate the
observed localization of deformation leading up to a rupture and subsequent delocalization of
deformation into the fractured medium afterward, as recently reported by \textcite{Kato2020,
BenZion2020}. This comprehensive representation of seismic activity underscores the importance of
considering fault volume effects in seismic simulations. While several existing models produce one,
or many, of the above-discussed features, the fault volume model stands out as the first to
reproduce the entirety of observations and statistics simultaneously, demonstrating its efficacy in
capturing the multifaceted nature of seismicity within fault systems.

We note here this is not the first attempt in our community to explain the diversity of slip
dynamics. In the late 1970's and early 1980's there was a flurry of papers by
\textcite[][]{Blandford1975, Nur1978, aki1979, andrews1981, hanks1981, Gusev1983} (among many
others) that have proposed heterogeneities in tractions, friction and material properties play a
major role in fault mechanics. Within this historical context, our work provides a natural framework
to introduce traction heterogeneities through geometric complexity and reproduces a broader spectrum
of slip dynamics than just earthquakes.

This fault volume model achieves all of the above without the explicit incorporation of frictional
heterogeneity or fluid presence within the medium, indicating its efficacy in explaining complex
slip behaviors solely through geometric complexity, which is an independently measurable quantity. A
notable distinction arises when comparing the behavior of a single fault with frictional
heterogeneity to that of a fault volume. In the former case, frictional properties remain constant
throughout seismic cycles, while with a fault volume model, traction heterogeneities (asperities or
barriers) naturally develop and disappear over various spatiotemporal scales. What may appear as a
barrier, region of high-coupling for example, during one seismic cycle (few decades) may become an
asperity during the next cycle, and vice-versa. This raises the question of what we mean by
``barriers'' or ``asperities'' in the context of geodetic observations, which typically span only a few decades. Such
observations are often interpreted as evidence of frictional asperities, yet in reality they simply
show that slip occurs when fault traction exceeds frictional resistance. Slip dynamics - whether
creep, slow slip events (SSEs), broadband earthquakes or earthquakes - are additionally governed by the temporal evolution
of tractions: they also control the nucleation length and thus determines whether the fault can
accelerate to seismic slip velocities. Moreover, no matter how good observations become, it will be
impossible to determine the frictional heterogeneity of a fault unless very strict model assumptions
are made. Therefore, we argue that modeling the full spectrum of slip dynamics through geometric
complexity offers a simpler, integrated, and comprehensive approach opening up a new avenue to
develop a digital twin framework for fault systems.

Recent advances in computational earthquake physics have enabled the development of digital twin
frameworks for fault systems. \textcite{Henneking2025} developed a Bayesian inversion-based digital
twin that couples physics-based models of earthquake rupture and tsunami generation with real-time
seafloor observations, demonstrating feasibility of real-time forecasting for subduction zone
events. Their approach combines 3D coupled acoustic-gravity wave equations with seafloor pressure
sensor data to infer earthquake-induced seafloor motion and forecast tsunami propagation with
quantified uncertainties. Complementing this operational focus, \textcite{Abdelmeguid2022, Mia2023,Peng2025}
have developed computational platforms for simulating sequences of earthquakes and aseismic slip
(SEAS) with explicit representation of fault zone complexity, including off-fault plasticity and
secondary fractures. Their hybrid finite element-spectral boundary integral method aims toward
``realizing Digital Twins for crustal volumes.'' These efforts demonstrate complementary approaches
to digital twins in earthquake science: real-time data assimilation for operational forecasting and
high-fidelity physics-based simulation of fault zone processes across multiple spatiotemporal
scales. 

We adopt the latter definition of a digital twin for our physics-based fault volume modeling
framework following the broader definition used in computational mechanics as ``a high accuracy
physics-based model'' validated against observations \parencite{Wright2020}. Our framework provides
a simple yet high-fidelity virtual representation reproducing the full spectrum of observed fault
slip behaviors from first principles, its statistical and scaling laws, enabling hypothesis testing
and scenario exploration impossible on natural fault systems. This digital twin generates synthetic
data with perfectly known sources and complete observational coverage, allowing us to develop
improved metrics and inverse methods for interpreting fault slip dynamics that we can then apply to
real observations. Future incorporation of data assimilation will enhance predictive capabilities,
but the current physics-based approach already establishes this framework as a powerful digital twin
for fault slip dynamics.

While our framework demonstrates the digital twin concept for fault slip dynamics, several avenues
merit further investigation. An inexhaustive list is compiled below. First, exploring a broader
range of friction parameters, and laws, for both off-fault fractures and the main fault would help
establish the generality of our findings across different geological settings. The power-law
exponent of the off-fault fracture size distribution represents another key parameter whose
systematic variation could reveal how fault network geometry controls the emergent spectrum of
seismic and aseismic phenomena along with the statistical properties. While we observed that
off-fault orientation influences aftershock duration—particularly when fractures align with the main
fault—a comprehensive study of orientation effects remains for future work. Additionally, our
current framework's exclusive focus on shear fractures can lead to unrealistic normal traction
accumulation; implementing a strategy to properly manage normal traction loading and unloading would
better capture the full traction evolution during slip events. To definitively confirm broadband
earthquakes, source time functions should be propagated through elastic waves to synthetic
seismometers. Moving toward fully dynamic simulations would eliminate quasi-dynamic approximations
and allow investigation of inertial effects, though at significantly higher computational cost.
Finally, the physical origin of the observed convexity in the truncated Gutenberg-Richter
distribution warrants theoretical investigation to understand whether it reflects fundamental
constraints on maximum earthquake size or emerges from specific properties of our fault network
geometry.

The model makes several testable predictions that can be evaluated using independent observations
not directly targeted in this study, including (i) systematic localization and migration of
microseismicity toward the eventual rupture plane prior to large events, (ii) threshold-dependent
scaling of slow-slip duration inferred from high-resolution geodetic or seismic catalogs, (iii)
characteristic spatiotemporal clustering of off-fault seismicity during slow slip, and (iv)
statistical properties of source time functions across scales. High-resolution seismicity catalogs,
dense nodal arrays, and improved geodetic inversions provide promising avenues to quantitatively
test these predictions and potentially falsify the model. Such evaluations will help refine the
fault volume framework and its applicability to real fault systems.

\section{Conclusion}

In this paper, we investigate how fault zone architecture can reproduce the observed spectrum of
slip dynamics and their statistical properties. We perform quasi-dynamic simulations on randomly
generated 2D fault zone models, where the fault zone consists of a main fault with self-similar
roughness and a power-law size distribution of off-fault fractures spanning lengths from about 1
meter to kilometers. The off-fault fracture density decays with distance from the main fault, over a
few cohesive zone length scales, following a power-law distribution. All faults are assigned
spatially uniform, slip rate-weakening frictional properties, and the critical slip distance $d_c$
is scaled with fault length such that each fracture, if simulated independently, can undergo seismic
cycles. We show that these ingredients are sufficient to reproduce the full spectrum of slip
dynamics, from slow slip events to fast earthquakes, and their statistical properties.

\begin{figure}
	\centering
	\includegraphics[width=1\textwidth]{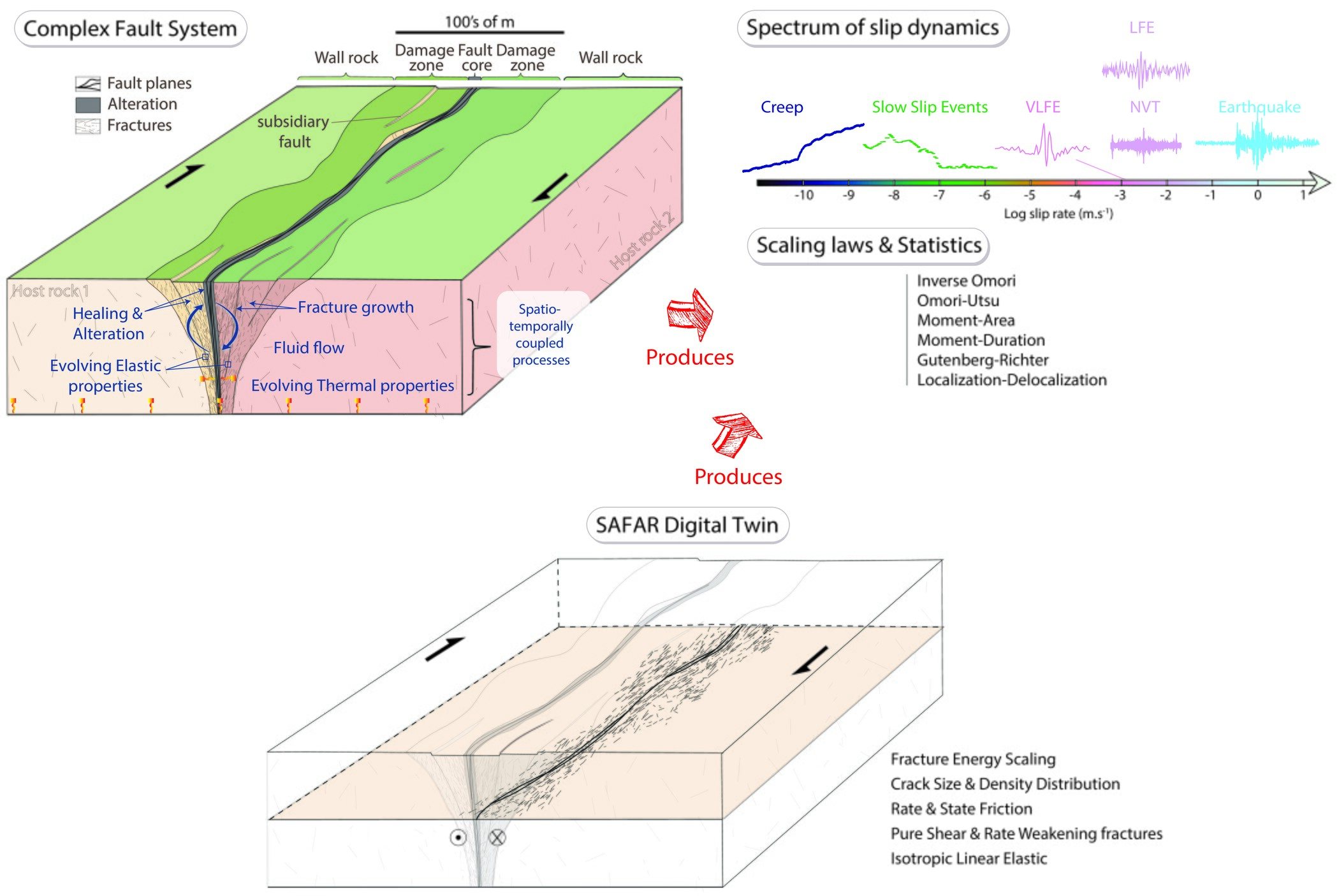}
	\caption{Summary of the digital twin model, SAFAR, that reproduces broadband slip dynamics,
	statistics and scaling laws of a fault volume.}
	\label{grandsummary}
\end{figure}

Our simulations reproduce all of the observed scaling and statistical properties of slip
dynamics, including the Gutenberg-Richter magnitude-frequency
distribution, Omori and inverse-Omori behavior of seismicity rates before and after mainshocks,
moment–duration and moment–area scaling relationships, and the observed localization-delocalization
of seismicity around an event. We also observe that off-fault seismicity migrates in a manner
similar to fluid diffusion in a homogeneous porous medium and raise a caveat emptor that
square-root-time-like migration can emerge from the physics of interacting faults governed by
rate-and-state friction, even in the absence of fluid diffusion.

Through an extensive analysis of the moment rate functions (MRFs), we find that off-fault MRF
durations are continuous, whereas MRFs on the main fault show a bimodal behavior—i.e., either slow
slip events or large fast earthquakes emerge. Additionally, the MRFs of small-magnitude slow-slip
events tend to be asymmetric, while those of fast-slip events display a more symmetric pattern. As
magnitude increases, the MRFs exhibit more pronounced acceleration and deceleration phases,
deviating from simple functional forms and revealing complexities such as multiple phases of moment
release.

We thus propose a geometry-driven modeling framework in which fault geometry provides a primary
organizational structure for slip dynamics. Fault-zone geometry is an independently measurable and
unavoidable property of natural systems across scales; in our framework, it naturally generates
spatiotemporally evolving traction heterogeneities that govern nucleation, rupture velocity, and
event interactions across the full spectrum of slip behaviors. This perspective helps reduce
reliance on poorly constrained parameters and provides a physically grounded reference framework. 

We do not intend to replace frictional or hydraulic heterogeneity models, but rather propose that
geometric complexity constitutes a necessary baseline upon which these processes operate. Frictional
and hydraulic heterogeneities may further modulate slip dynamics, but they act within a fault system
whose stress organization is primarily shaped by geometry. This geometry-driven baseline provides a
foundation within which additional complexities can be meaningfully evaluated and their relative
contributions assessed. Such a framework enables the construction of digital twins of real fault
systems—computational analogs that capture essential geometric and mechanical properties while
generating realistic slip dynamics.

The fault volume model presented in this paper demonstrates the viability of this approach. Despite
its simplicity and reliance on broadly accepted assumptions, it successfully reproduces many of the
observed key physical and statistical features of fault zones (see Figure \ref{grandsummary}). We
name this digital twin framework \textit{SAFAR} (Système Analogue des FAilles Réelles). Our
synthetic catalogs can enhance machine learning algorithms for seismic hazard forecasting by
providing physically consistent training data that spans the full spectrum of seismic behaviors. The
framework also provides data to analytically model various physical mechanisms that underlie the
observed scaling laws, statistics, and slip dynamics. In addition, it can be used to characterize
the response of a fault system to various types of perturbations, such as fluid injection or earth
tides, and to understand the underlying physics of earthquake triggering. By demonstrating that
geometric complexity alone can generate rich fault dynamics without requiring unmeasurable
frictional heterogeneities, this work provides a novel framework for more predictive and physically
grounded seismic hazard assessment.

\section*{Acknowledgements}

MA, NK, CV, JC, AG and HSB gratefully acknowledge the European Research Council (ERC) for its full
support of this work through the PERSISMO grant (No. 865411). MYT acknowledges support from the
Agence National de la Recherche (ANR) IDEAS contract ANR-19-CE31-0004-01. PR acknowledges support
from the GPX program,funded by the Agence National de la Recherche (ANR), CGG, TOTAL, and
Schlumberger for his PhD fellowship and the European Research Council (ERC) Starting Grant 101040600
(HYQUAKE). NK acknowledges funding from Horizon Europe (ChEESE-2P project, grant agreement No.
101093038) for partial support in working on this manuscript. This paper has benefited from kind
discussions and feedbacks from Satoshi Ide, Hideo Aochi, Romain Jolivet and Raul Madariaga. The
authors also acknowledge the feedback from the editor R. Abercrombie, the associate editor A-A.
Gabriel, Prof. Y. Huang and an anonymous reviewer. The numerical simulations presented in this study
were performed on the MADARIAGA cluster, also supported by the ERC PERSISMO grant. We also used LLM
models like GPT-4 and Claude to debug and optimize our codes.


\section*{Conflict of Interest}
The authors declare no conflicts of interest relevant to this study.
\section*{Open Research Section}
The datasets generated and analyzed during this study, and codes used to analyze, are available at \textcite{kheirdast2026}.

\clearpage


\begin{center}

\LARGE {\bf Supplementary Information for\\[6pt]}
\LARGE {\bf ``Fault volume digital twin to reproduce the full slip spectrum, scaling and statistical
laws'' \\[12pt]}

\normalsize
M. Almakari$^{1,\dagger}$, N. Kheirdast$^{1,\dagger}$, C. Villafuerte$^{1,*}$, M. Y. Thomas$^{2}$,
P. Dubernet$^{1}$, J. Cheng$^{1,\S}$, A. Gupta$^{1}$, P. Romanet$^{3,4}$, S. Chaillat$^{5}$, H. S.
Bhat$^{1,\P}$ \\[12pt]

\begin{enumerate}
	\scriptsize
	\setlength\itemsep{-5pt}
	\item Laboratoire de Géologie, Ecole Normale Superieure, CNRS-UMR 8538, PSL Research University,
	Paris, France\\
	\item Université de Rennes, CNRS, Géosciences Rennes, CNRS-UMR 6118, Rennes
	\item Department of Earth Sciences, La Sapienza University of Rome, Piazzale Aldo Moro 5, 00185
	Roma, Italy
	\item Université Côte d`Azur, CNRS, IRD, Observatoire de la Côte d'Azur, Géoazur,
	Sophia-Antipolis, 06560 Valbonne, France
	\item Laboratoire POEMS, CNRS-INRIA-ENSTA Paris, Institut Polytechnique de Paris
\end{enumerate}

\let\thefootnote\relax\footnotetext{$\dagger$ M. Almakari and N. Kheirdast contributed equally to this work. 
* Currently at Instituto de Geofisica, Universidad Nacional Autonoma de México. 
$\S$ Currently at Division of Geological and Planetary Sciences, California Institute of Technology. 
$\P$ Corresponding author: \texttt{harshasbhat@gmail.com}}

\end{center}


\section*{Introduction}

This Supplementary Information provides detailed technical documentation for the numerical methods, algorithms, and additional results supporting the main manuscript. The document is organized as follows:

\textbf{Section S1} describes the quasi-dynamic earthquake cycle model

\textbf{Section S2} presents the catalog building algorithm used to identify and characterize individual slip events from spatiotemporal slip rate matrices. 

\textbf{Supplementary Figures:}

\textbf{Figure S1} illustrates the catalog generation workflow.

\textbf{Figure S2} demonstrates spatial segregation of fast and slow events in a fault system with off-fault fractures parallel to the main fault, using the same configuration as the main manuscript.

\textbf{Figure S3} presents statistical properties and scaling laws

\textbf{Figure S4} shows moment-duration scaling relationships with different detection thresholds.

\clearpage
\setcounter{section}{0}
\renewcommand\thesection{S\arabic{section}}
\renewcommand\thesubsection{S\arabic{section}.\arabic{subsection}}
\renewcommand{\theequation}{S\arabic{section}.\arabic{equation}}
\setcounter{equation}{0}
\renewcommand{\thefigure}{S\arabic{figure}}
\setcounter{figure}{0}

\renewcommand{\theHsection}{appendix.\Alph{section}}
\renewcommand{\theHsubsection}{appendix.\Alph{section}.\arabic{subsection}}
\renewcommand{\theHequation}{appendix.\Alph{section}.\arabic{equation}}

\section{Model of earthquake cycle}
\label{QDBIEMmodel}

\subsection{Governing equation}

A classic model of earthquake cycles contains: (1) an elastic medium that stores strain energy, (2)
a physical mechanism that loads the system (plate tectonics), and finally (3) a frictional
resistance on the fault, that allows the accumulation of strain on the fault. To build the model of
earthquake cycles in this paper, we will use the linear momentum balance equation that states that
the stress loading on the fault, and the elastic stress response due to a slip distribution on the
fault, must be equal to the strength of this fault at each curvilinear point $s$ along the fault.
Mathematically, it is written as:
\begin{equation}
	\tau^f(s) = \tau_t^{\text{el}}(s) + \tau^{\text{rad}}(s)+\tau^{\text{load}}(s)
	\label{momentum_balance}
\end{equation}
Where $\tau^f(s)$ is the strength (model by a friction law) of the fault, $\tau_t^{\text{el}}(s)$ is
the elastostatic tangential traction response due to the slip distribution on the fault,
$\tau^{\text{rad}}(s)$ is the instantaneous response of the system that allows some inertia control
\parencite{Rice1993}, these two latter modeling the reaction of the elastic medium due to the slip.
Finally, $\tau^{\text{load}}(s)$ is the loading shear traction on the fault. Each of these terms
will be described with more detailed in the following sections. 

\subsection{Strength of the fault}
\label{frictionlawappendix}
We will model the strength of the fault by the regularized rate-and-state friction law
\parencite{Lapusta2000}: 

\begin{equation}\label{eq:regularized_rate_and_state}
	\tau^f(s,t) = -f(V,\theta)\sigma_n(s,t) = -a\sigma_n(s,t) \sinh^{-1}\left[\frac{V(s,t)}{2V_0} \exp\left\{\frac{f_0 + b\ln[V_0\theta(s,t)/d_c]}{a}\right\}\right]
\end{equation}

with aging state evolution \parencite{Dieterich1979,Ruina1983}:

\begin{equation}
	\dfrac{d\theta(s,t)}{dt} = 1 - \frac{\theta(s,t) V(s,t)}{d_c}
\label{state_evolution}
\end{equation}

$a$ is the direct effect parameter that governs the instantaneous change in friction with a change
in slip rate, $b$ is the evolution effect parameter which controls how friction evolves over
time via changes in the state variable, $d_{c}$ is the characteristic slip for state evolution,
$V_0$ is reference slip rate and $f_{0}$ is the reference friction at $V_0$. $V$ and $\theta$
represent respectively the slip velocity and the state variable. $\sigma_n$ is the normal traction
on the fault and $d_c$ is the critical slip distance. The term $f_0 + b \log (  V_0 \theta / d_c)$
is sometimes denoted by $\Psi$. 

\subsection{Elastostatic response of the fault due to a slip distribution}

The normal traction and tangential traction on the fault, given a slip distribution
$\Delta\mathbf{u}(s)$, and the fault local curvature $\kappa^t(s)$, in an infinite and homogeneous
medium can be calculated using boundary element method \parencite{Tada1997,Romanet2020,Romanet2024}.
For convenience we will note the tangential slip on the fault $\Delta u_t(s) = \mathbf{t}(s)\cdot
\Delta\mathbf{u}(s)$:

\begin{equation} 
	\tau_{t}^{\text{el}}(s) = \mathbf{t}(s) \cdot \overline{\overline{\sigma}}(\mathbf{y}(s)) \cdot \mathbf{n}(s) = \int_{\Gamma} \! K_{\text{grad}}^{t}(s,\xi) \frac{\partial}{\partial \xi}\Delta u_{t}(\xi) \mathrm{d}\xi +\int_{\Gamma} \! K_{\text{curv}}^{t}(s,\xi) \kappa^t(\xi)\Delta u_{t}(\xi) \mathrm{d}\xi
	\label{tau_t}
\end{equation}

\begin{equation} 
	\sigma_{n}^{\text{el}}(s) = \mathbf{n}(s) \cdot \overline{\overline{\sigma}}(\mathbf{y}(s)) \cdot \mathbf{n}(s) = \int_{\Gamma} \! K_{\text{grad}}^{n}(s,\xi) \frac{\partial}{\partial \xi}\Delta u_{t}(\xi)  \mathrm{d}\xi	+\int_{\Gamma} \! K_{\text{curv}}^{n}(s,\xi) \kappa^t(\xi)\Delta u_{t}(\xi) \mathrm{d}\xi
	\label{tau_n}
\end{equation}

In order to set the model of earthquake cycle, we chose the compression negative sign convention.
$s$ is the curvilinear location along the fault, $\Gamma \equiv \mathbf{y}(s)$, at which the stress
is evaluated. $\xi$ is a curvilinear location, at which a source of stress is located.
$\mathbf{n}(\xi)$ and $\mathbf{t}(\xi)$ are respectively the normal and tangential vector to the
fault at point $\xi$. Their component in the global coordinate system are written with the
corresponding subscript. Finally, $K^{t}_{\text{grad}}$ is the kernel for the tangential traction,
and $K^{n}_{\text{grad}}$ and $K^{n}_{\text{curv}}$ are the kernel for the normal traction.
$K^{t}_{\text{grad}}$ and $K^{t}_{\text{curv}}$ are the kernels for the tangential traction
associated respectively to the effect of the gradient of tangential slip and local curvature that
multiplies the tangential slip. They both derive from the fact that a derivative of the tangential
slip vector along the fault is:
\begin{equation}
\begin{split}
\frac{\partial}{\partial \xi}[\Delta u_{t}(\xi)\mathbf{t}(\xi)]&= \frac{\partial}{\partial \xi}[\Delta u_{t}(\xi)]\mathbf{t}(\xi)+\Delta u_t(\xi)\frac{\partial}{\partial \xi}\mathbf{t}(\xi) \\ 
&=\frac{\partial}{\partial \xi}[\Delta u_{t}(\xi) ]\mathbf{t}(\xi)+\Delta u_t(\xi)\kappa^t(\xi)\mathbf{n}(\xi)
\end{split}
\end{equation}
where we have used the relationship $\frac{\partial}{\partial \xi}\mathbf{t}(\xi) =
\kappa_t(\xi)\mathbf{n}$.

Similarly, $K^{n}_{\text{grad}}$ and $K^{n}_{\text{curv}}$ are the kernels for the normal traction.
Mode details about the derivation of these equations can be found in \textcite{Romanet2024}. It
should be noted here that these kernels take into account the full elastic interaction between all
faults in the system, including the main fault and all off-fault fractures.

\subsection{Radiation damping term}

The radiation damping term was first introduced by \textcite{Rice1993} in the context of
quasidynamic modeling. In fact \textcite{Andrews1980} showed that this is the impedance of the fault
(ratio of shear stress to slip velocity) in the long wavelength limit. It was later shown that this
term is exactly accounting for the instantaneous shear stress drop of the fault due to sliding
\parencite{Cochard1994} : 

\begin{equation}
	\tau^{\text{rad}}(s) = - \dfrac{\mu}{2c_s}V(s) 
\end{equation} 

where $c_s$ is the shear wave speed. We use this term together with the static kernel to account for
some dynamics in the system. Without this term, the slip on the fault during one event would be
unbounded \parencite{Rice1993}.

\subsection{Set of ordinary differential equations for Quasi-Dynamic earthquake cycle models}

Balance of forces requires the strength of the fault to be equal to the elastic shear traction (due
to slip) plus the far-field loading traction plus radiation damping term (eq.
\ref{momentum_balance}). 

By differentiating eq. \ref{momentum_balance} with time $t$, it can be recast into a set of coupled
ODEs. The slip acceleration becomes:

\begin{equation}
	\dfrac{dV(s,t)}{dt} = \frac{\dot{\tau}^{\text{load}}(s,t)  +\dot{\tau}_t^{\text{el}}(s,t)+ \dfrac{\partial f(V,\theta)}{\partial \theta}\dfrac{d\theta}{dt} \sigma_n(s,t)   + f(V,\theta) \dot{\sigma}_n(s,t)}{\dfrac{\mu}{2c_s} - \dfrac{\partial f(V,\theta)}{\partial V}\sigma_n(s,t)  }
	\label{ode1}
\end{equation}

From this equation, it is easy to see that the denominator would go to zero and hence the
acceleration would go to infinity if there was no radiation damping term \parencite{Rice1993}.

The elastic shear traction rate is given by:
\begin{equation}
	\dot{\tau}_t^{\text{el}}(s,t) =\int_{\Gamma} \! K_{\text{grad}}^{t}(s,\xi) \frac{\partial}{\partial \xi} V(\xi,t) \mathrm{d}\xi +\int_{\Gamma} \! K_{\text{curv}}^{t}(s,\xi) \kappa^t(\xi)V(\xi,t) \mathrm{d}\xi
\end{equation}

The change of normal traction has two sources: one due from the loading rate and another one due to
the reaction to slip of the elastic medium:

\begin{equation}
	\dfrac{d\sigma_n(s,t)}{dt} = \int_{faults}K_{\text{grad}}^{\text{n}}(s,\xi)\dfrac{\partial}{\partial \xi} V (\xi,t)\mathrm{d}\xi +\int_{faults}K_{\text{curv}}^{\text{n}}(s,\xi) \kappa^{\text{t}}(\xi) V (\xi,t)\mathrm{d}\xi+ \dot{\sigma}_n^{load}(s,t)
\end{equation}

And we recall the evolution of the state variable eq. \ref{state_evolution}. This set of ODEs is
then solved at each centre of element, using the Runge-Kutta 45 adaptive time step ODE solver
algorithm \parencite{cash1990}.

\subsection{Numerical discretisation of the Boundary Integral Equation}

In order to evaluate the previous singular integrals in the sense of Cauchy principal values, we
will assume piece-wise constant slip over fixed length $\Delta s$, centred on $\mathbf{y}(s_i)$. The
slip is discretised as follows \parencite{Rice1993,Cochard1994}:
\begin{equation}
	\Delta u(s) = \sum_{j=1}^N \Delta u(s_j) [\mathcal H(s-s_j+\Delta s/2)-\mathcal H(s-s_j -\Delta s/2)]
\end{equation}
where $\mathcal H$ is the Heaviside function and $N$ is the number of elements used to discretise
the fault. Because the kernels imply the evalutation of the tangential vector, it is also needed to
discretised it along the fault:
\begin{equation}
	\mathbf{t}(s) = \sum_{j=1}^N \mathbf{t}(s_j) [\mathcal H(s-s_j+\Delta s/2)-\mathcal H(s-s_j -\Delta s/2)]
\end{equation}

Then the boundary integral equations \ref{tau_t} and \ref{tau_n} become a summation:
\begin{eqnarray} 
	\label{discreteBEM}
	\tau(s_i) = \sum_{j=1}^{N}  [K_t(s_i,\xi_{j})-K_t(s_i,\xi_{j+1})] \Delta u_j \nonumber \\ 
	\sigma(s_i) = \sum_{j=1}^{N}  [K_n(s_i,\xi_{j})-K_n(s_i,\xi_{j+1})] \Delta u_j  
\end{eqnarray}

The implementation of this formulation requires to calculate for each $\tau(s_i)$ ($N$ terms) a sum
over $N$ terms (for each $j$), which leads to a computational complexity of $\mathcal O(N^2)$. It
means that the computational time will grow with the square of the problem size. This makes it
difficult to handle large problems with a straightforward implementation. However, for the majority
of the cases, the kernel $K$ is smooth when the source point $\mathbf{y}(\xi_j)$ is far enough from
the evaluated point $\mathbf{y}(s_i)$, or in other words:$|\mathbf{y}(s_i)-\mathbf{y}(\xi_j)|>>1$.
Several methods can be used to accelerate the evaluation of eqns. \ref{tau_t} and \ref{tau_n}. The
Fast Fourier Transform was the first method used to accelerate the calculation
\parencite{Andrews1985}, however it it requires equispaced points on the fault. The only
configuration where this can be achieved is on a simple planar fault. Fast fourier transform has
been widely used in numerical models of seismic cycles
\parencite{Lapusta2000,Lapusta2009,Chen2009,Michel2017}. The Fast Multipole Method allows for
considering a complex geometry, and was used by some numerical models
\parencite{Hirahara2009,Romanet2018}, however it requires an analytical development of the kernel.
In this paper, we employ Hierarchical matrices which allows for an algebraic development of the
kernel and thus can be applied to a large variety of problems. This approach is particularly optimal
for static problems. This will be discussed in the next section. 

\subsection{Hierarchical Matrices}

Hierarchical matrices have already been used in the context of quasi dynamic modeling of faults
\parencite{Bradley2014, romanet2017b, Cheng2025}. Hierarchical matrices provide data-sparse
approximations of non-sparse (dense) matrices \parencite{Hackbusch1999, Borm2003}. Hierarchical
matrices provide an approximation requiring only $\mathcal{O}(N \log{N})$ units of storage (instead
of $\mathcal O(N^2)$). The two advantages of using H-matrices are that it reduces the memory for
saving the full matrix of interaction, and it is also reducing the number of operations to perform
the matrix-vector products. The use of H matrices is extremely pertinent in quasidynamic models. The
kernel matrix of stress interaction \ref{discreteBEM} is not changing over time, making it a
suitable matrix for compression, since the most expensive part, the construction of the data-sparse
representation, is performed only once, while subsequent matrix-vector products are extremely fast.
The accuracy of this method is completely controlled by the tolerance set for the compression,
particularly if the required minimum accuracy approaches machine precision. The mathematical
formulation of hierarchical matrices is beyond the scope of this paper but is very well explained in
\textcite{Borm2003, Desiderio2017, romanet2017b, Cheng2025}. 

\subsection{Frictional length scales}
\label{frictionallengthscales}
Although the general answer is unavailable in the general framework of rate and state, it is
possible to infer a length scale in some limiting cases for simple straight continuum faults with
homogeneous frictional parameters. The first nucleation length-scale derived for continuous fault
came from the spring slider modeling. If it is assumed that the stiffness of a fault $k_{fault}$ is
inversely proportional to its length, we can derive a nucleation length scale $L_{nuc}$
\parencite{rice1992}. At the critical length $L_{nuc}$, the stiffness of the fault equals the
critical stiffness derived from a spring slider \parencite{rice1992}.
\begin{equation}
k_c = \sigma_n \frac{b-a}{d_c} = \frac{\mu}{L_{nuc}} = k_{fault}
\end{equation}
Thus,
\begin{equation}
L_{nuc} = \frac{\mu d_c}{\sigma_n(b-a)}
\end{equation}
By studying the nucleation on fault with rate and state resistance, Dieterich derived another
nucleation length scale $L_b$ inversely proportional to the friction parameter $b$
\parencite{dieterich1992}. 
\begin{equation}
L_{b} = \frac{\mu d_c}{\sigma_n b}
\end{equation}
More recent work shows that the nucleation actually depends on the ratio $a/b$
\parencite{Rubin2005,ampuero2008a,viesca2016a}. \textcite{Rubin2005} first derived analytical
solution in the limiting case where ${V \theta}/{D_c}>>1$ for the aging state evolution law. They
showed that this assumption would remain valid only if the ratio $a/b< 0.3781$.  For high $a/b$
however, they pointed out that the coefficient ${V \theta}/{D_c}$ was nearly constant at the
interior of the nucleation patch. Using that as an assumption, with energetic consideration, they
were able to derive another expression for the nucleation length scale when $a/b$ approaches $1$. We
can summarize their result by:
\begin{equation}
 \begin{cases} 
      L_{nuc}= 2\times 1.3774L_b                    & 0\le a/b< 0.3781 \\
      L_{nuc}= 2\times\dfrac{L_b}{\pi(1-a/b)^{2}}    &  a/b \to 1
   \end{cases}
   \label{L_nuc}
 \end{equation}
 \newline
 The nucleation length scale found by \textcite{Rubin2005}, was later shown to hold true by
 recasting the system of equations to look for instabilities and doing a linear perturbation
 analysis of these instabilities \parencite{viesca2016a}.

\section{Catalog building algorithm}\label{catalog_algorithm}

\begin{algorithm}
	\caption{Identify events}
	\begin{algorithmic}[1]
		\Procedure{ConnectedComponentsWithProperties}{image, rows, cols} \State // \texttt{image} :
		binary matrix where 1 indicates active pixel, 0 indicates background \State //
		\texttt{labels} : integer matrix where each nonzero entry is the region label \State //
		\texttt{properties} : array of structures with min/max row/column indices per region \State
		Initialize \texttt{labels[rows, cols]} $\gets 0$ \State $current\_label \gets 0$ \State
		Define offsets: $dr \gets [-1, -1, -1, 0, 0, 1, 1, 1],\ dc \gets [-1, 0, 1, -1, 1, -1, 0,
		1]$ \For{$col \gets 1$ to $cols$} \For{$row \gets 1$ to $rows$} \If{$image[row, col] = 1$
		and $labels[row, col] = 0$} \State $current\_label \gets current\_label + 1$ \State
		$properties[current\_label]$: min/max row/col $\gets row$, $col$ \State
		\Call{LabelRegions}{$row, col, current\_label$} \EndIf \EndFor \EndFor \State \Return
		$labels$, $properties$, $current\_label$ \EndProcedure \Procedure{LabelRegions}{$r, c,
		label$} \If{$r$ or $c$ out of bounds or not 1 or already labeled} \Return \EndIf \State
		$labels[r,c] \gets label$ \State Update $properties[label]$ with min/max of $r$, $c$ \For{$i
		\gets 1$ to $8$} \State $nr \gets r + dr[i]$, $nc \gets c + dc[i]$ \State
		\Call{LabelRegions}{$nr, nc, label$} \EndFor \EndProcedure
	\end{algorithmic}
	\label{algo}
\end{algorithm}

To identify events that are spatiotemporally contiguous we use a method called Connected Component
Labeling with Bounding Box Extraction. Several widely used software packages support connected
component labeling (CCL) and region property extraction. In MATLAB, the \texttt{bwlabel} and
\texttt{regionprops} functions provide robust tools for labeling and extracting geometric features
such as bounding boxes and centroids \parencite{mathworks2024}. In Python, the \texttt{scikit-image}
library offers similar functionality \parencite{vanderwalt2014}. OpenCV, a popular C++ and Python
computer vision library, includes highly optimized functions  
for fast labeling and region statistics \parencite{bradski2000}. In this work we decided to
implement this ourselves as the algorithm is trivial and it's easier to adapt the results of our
code to this algorithm (see Algorithm \ref{algo}).

\begin{figure}
	\centering
	\includegraphics[width=1\textwidth]{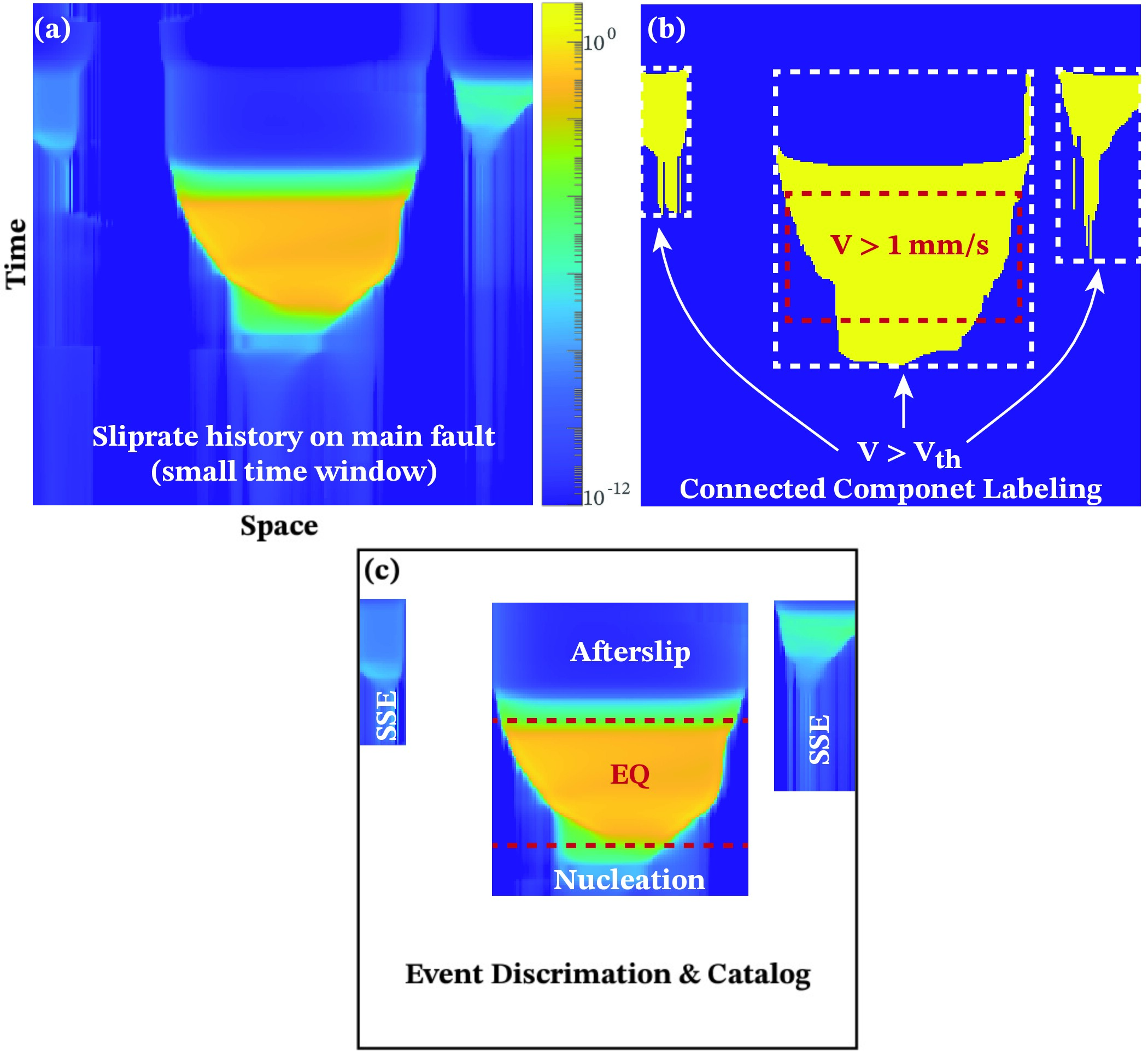}
	\caption{Steps involved in catalog generation from slip rate matrix. (a) Slip rate matrix (time,
	space). (b) Binary image after thresholding slip rates with Labeled connected components. (c)
	Bounding boxes around each connected component, representing individual events.}
	\label{catalogexplained}
\end{figure}

We first take a slip rate matrix (time, space) and convert into a binary image by setting to 0 all
values that are below a velocity threshold. The algorithm then performs connected component analysis
on this binary image using a recursive depth-first search (DFS) strategy to identify and label
contiguous regions of foreground pixels (i.e., pixels with value 1). Each connected region is
assigned a unique integer label, and its spatial extent is recorded as a bounding box defined by the
minimum and maximum row and column indices. This type of region labeling is a fundamental technique
in image analysis, commonly used for shape recognition, segmentation, and morphological operations.
The DFS uses an 8-connectivity scheme, meaning that each pixel is connected to its horizontal,
vertical, and diagonal neighbors. This is particularly useful in natural or irregular structures
where connectivity extends beyond 4-neighbor (Manhattan) adjacency. The use of recursive DFS
provides an intuitive implementation but is best suited for small to moderately sized images due to
potential stack overflow risks. For larger datasets, stack-based or union-find methods may be
preferred \parencite{rosenfeld1966,shapiro2001}.

The procedure begins by scanning the image in a raster order. When an unlabeled foreground pixel is
encountered, a new label is assigned, and a recursive DFS is initiated from that pixel to visit all
8-connected neighboring pixels belonging to the same region. During traversal, the algorithm updates
a property array that maintains the axis-aligned bounding box for each region in the form
(\text{min\_row}, \text{max\_row}, \text{min\_col}, \text{max\_col}). Once the beginning and the end
of the event (\text{min\_row}, \text{max\_row}) and its spatial extent (\text{min\_col},
\text{max\_col}) are obtained its trivial to compute the Moment, average slip, duration and other
catalog based quantities.

\begin{figure}
    \includegraphics[width=\textwidth]{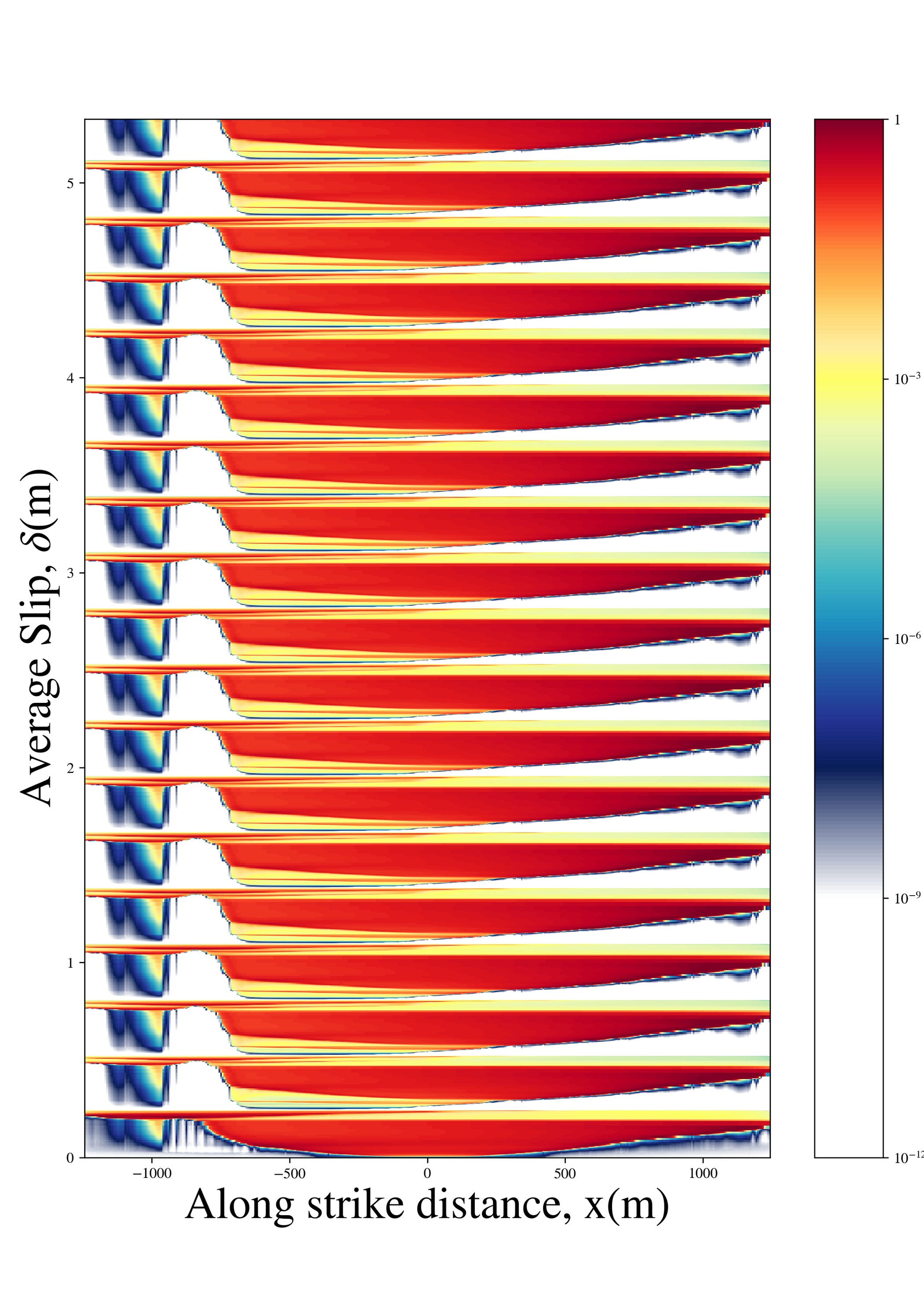}
    \caption{Segregation of slow and fast events at two separated regions of the fault. The geometrical configuration is the same as in Figure \ref{geometry}a except that the off-fault fractures are parallel to the main fault.}
    \label{segregation}
\end{figure}


\begin{figure}[!ht]
	\centering
	\includegraphics[width=\textwidth,]{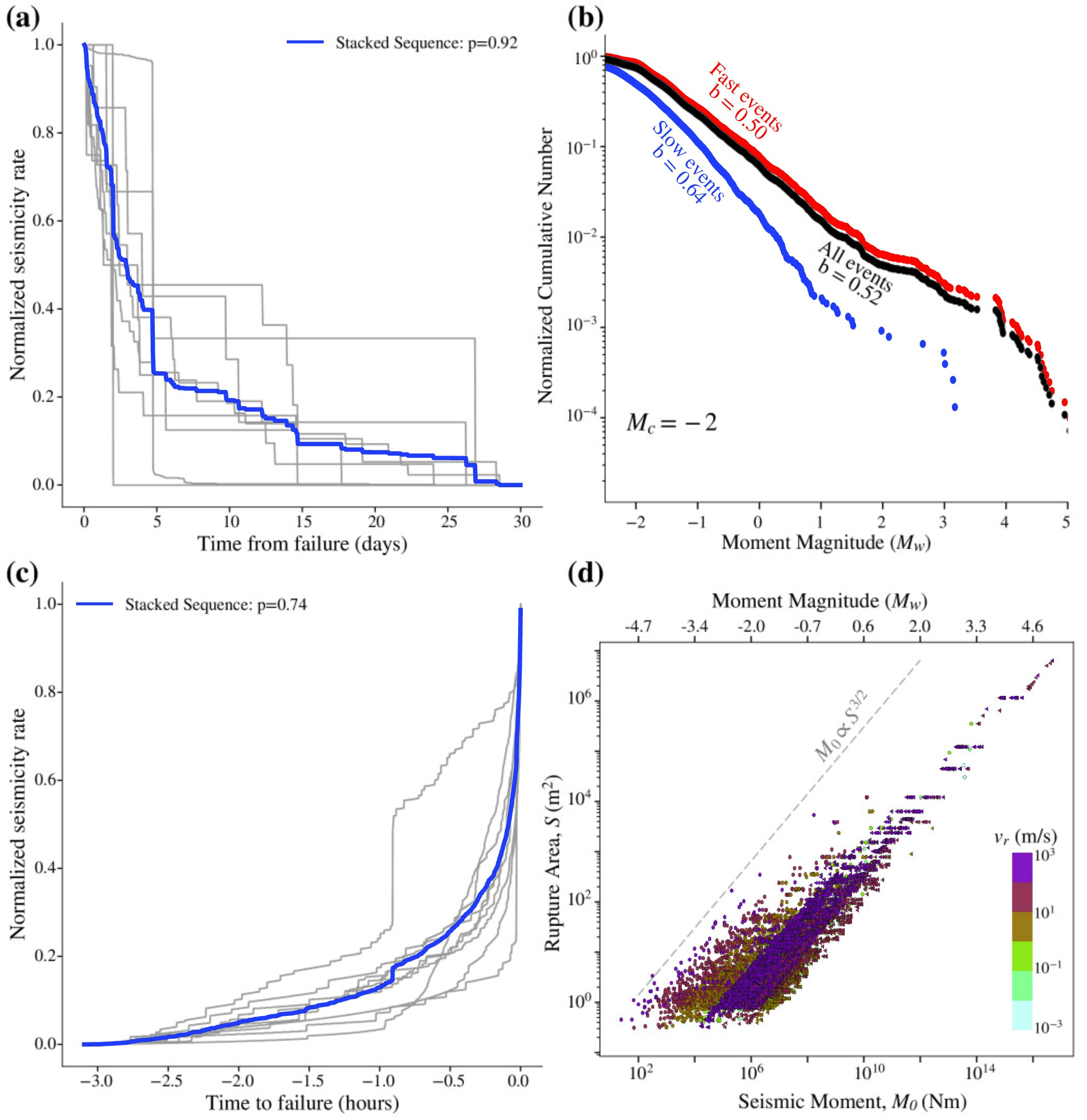}
	\caption{(a) Omori Law: Seismicity rate decreases over time following the mainshock. Each gray
	curve represents one earthquake cycle with one mainshock on the main fault, the blue curve
	represents the stacked sequences of all the seismic cycles in this simulation case (b)
	Magnitude-frequency distribution of the cataloged fast ruptures follows the Gutenberg-Richter
	distribution. Red and blue colors represent fast and slow ruptures respectively, while black
	color represents the total catalog (c) Inverse Omori: Seismicity rate increases inversely with
	time as the main rupture approaches. Each gray curve represents one earthquake cycle with one
	mainshock on the main fault, the blue curve represents the stacked sequences of all the seismic
	cycles in this simulation case (d) Scaling laws of inferred seismicity: Moment area scaling.
	Events are color-coded based on rupture velocity $v_r$.}
	\label{stats}
\end{figure}

\begin{figure}[!ht]
	\centering
	\includegraphics[width=\textwidth]{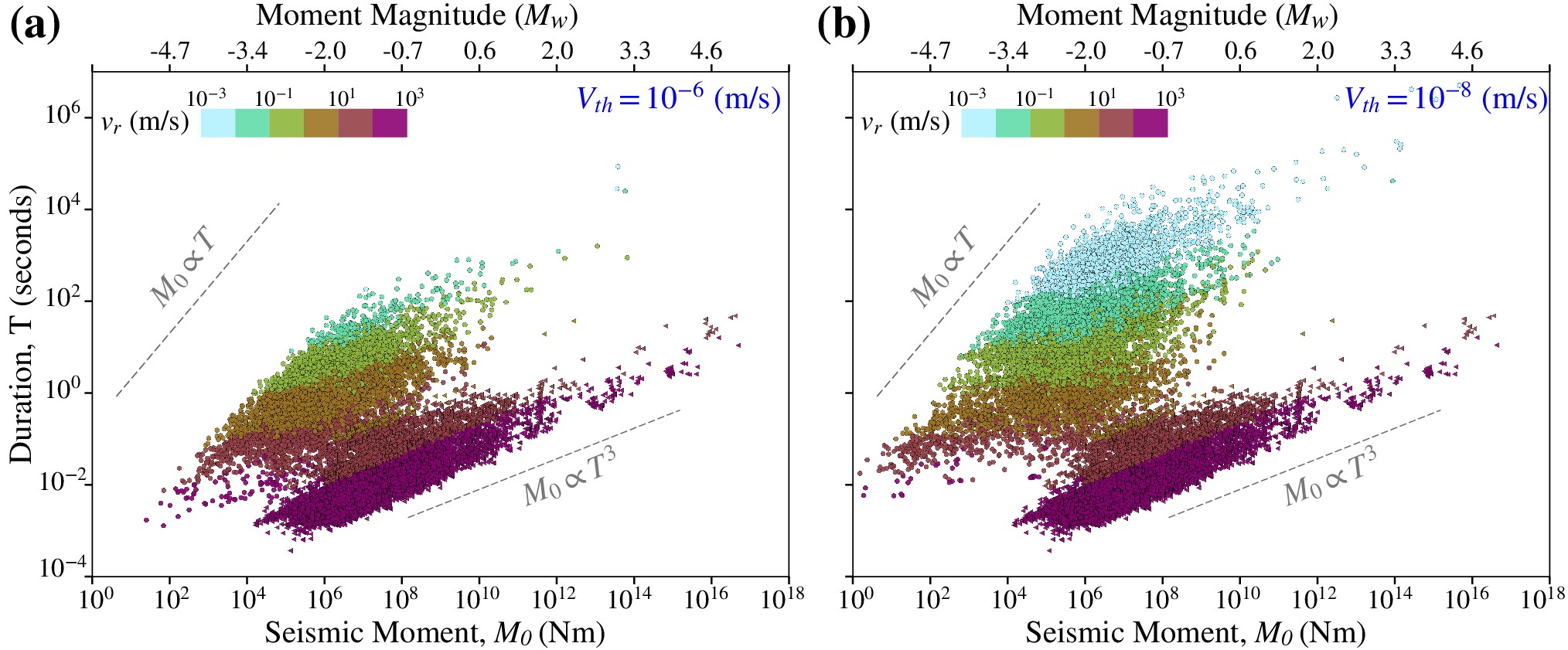}
	\caption{Scaling laws of inferred seismicity: Moment-duration scaling with a threshold of
	detection of slow ruptures at (a) $10^{-6}$ m/s and (b) $10^{-8}$ m/s. Events are color-coded
	based on rupture velocity $v_r$. }
	\label{scalingMT}
\end{figure}

\clearpage
\renewcommand*{\bibfont}{\scriptsize}
\printbibliography

\end{document}